\documentclass[aps,pra,preprint,pra,showpacs,amsmath,amssymb]{revtex4}

\usepackage{graphicx}
\usepackage{dcolumn}
\usepackage{bm}

\begin{document}
\preprint{submitted to Phys. Rev. A}
\author{H. Bergeron}
\affiliation{LURE bat. 209D Centre Universitaire Paris-Sud -BP34- 91898 
Orsay cedex, FRANCE}
\email{herve.bergeron@lure.u-psud.fr}
\title{New derivation of quantum equations\\from classical stochastic arguments}
\date{\today}

\begin{abstract}
In a previous article [H. Bergeron, J. Math. Phys. \textbf{42, } 3983 
(2001)], we presented a method to obtain a continuous transition from 
classical to quantum mechanics starting from the usual phase space
formulation of classical mechanics. This procedure was based on a
Koopman-von Neumann approach where classical equations are reformulated
into a quantumlike form. In this article, we develop a different derivation
of quantum equations, based on purely classical stochastic arguments, taking
some technical elements from the Bohm-F\'{e}nyes-Nelson approach. This study
starts from a remark already noticed by different authors [M. Born, 
\emph{Physics in My Generation} (Pergamon Press, London, 1956); E.
Prugove\v {c}ki, \emph{Stochastic Quantum Mechanics and Quantum Spacetime}
(Reidel, Dordrecht, 1986)], suggesting that physical continuous observables
are stochastic by nature. Following this idea, we study how intrinsic 
stochastic properties can be introduced into the framework of classical
mechanics. Then we analyze how the quantum theory can emerge from this
modified classical framework. This approach allows us to show that the
transition from classical to quantum formalism (for a spinless particle)
does not require real postulates, but rather soft generalizations.
\end{abstract}

\maketitle

\section{INTRODUCTION}

Since the beginning of quantum mechanics, different links have been
developed between classical and quantum frameworks, in order to overcome the
main difficulties due to the difference of formalisms, and to get a better
understanding of their interplay. These attempts can be roughly divided into
three families.

One family is represented by the Wigner-Weyl (WW) approach \cite
{wignerweyl1,wignerweyl2,wignerweyl3,wignerweyl4,wignerweyl5,wignerweyl6},
based on the reformulation of quantum mechanics into phase space thanks to a
continuous map. Particularly, the Wigner-Weyl transformation allows one to
recover the semi-classical limit of quantum mechanics. Moreover quantum
mechanics appears as a deformation of the Abelian function algebra in
phase-space where the standard multiplication is replaced by the so-called
$*_{h}$-product \cite{starproduct1,starproduct2,starproduct3} 
(deformation quantization). But the
Wigner-Weyl transformation is not the unique way to obtain such a continuous
map. The use of coherent states \cite
{coherentstates1,coherentstates2,coherentstates3,coherentstates4,coherentstatesberg}
(specially the coherent states of the Galilei group \cite{coherentprugo1})
allows one to obtain the same kind of correspondence.

Another family is represented by the Koopman-von Neumann (KvN) approach \cite
{kvn1,kvn2,kvn3,kvn4,kvn5,kvn6}, based on a reformulation of classical
mechanics into the Hilbert space language. This leads to a quantumlike
theory (but always classical) that can be directly compared to quantum
mechanics. Particularly, we can study how this quantumlike framework must be
modified in order to rebuild a true quantum theory \cite{kvnberg}.

These two attempts are in fact two sides of the same problem, namely finding
a unified mathematical framework for classical and quantum mechanics. Among
the numerous publications on this subject, and in addition to the authors
already mentioned, we can quote the works of G. Mackey \cite
{unifiedformalism1,unifiedformalism2} and E. Prugove\v {c}ki \cite
{unifiedprugo1,unifiedprugo2}.

The third attempt is very different, and it is represented by the
Bohm-F\'{e}nyes-Nelson (BFN) approach \cite
{bfn1,bfn2,bfn3,bfn4,bfn5,bfn6,bfn7,bfn8,bfn9}. Above all, this approach is
centered on the problem of the interpretation of quantum mechanics. The
leading idea is that quantum phenomena are due to some stochastic effects
that take place on a classical background where the idea of trajectory
(obtained from equations of motion) remain valid. The Planck constant 
``$\hslash $'' becomes a ``measure'' of the stochastic effects. 
Although this interpretation can lead to disagreements with quantum
mechanics, especially in the case of non-interacting subsystems
\cite{bfn10a,bfn10b,bfn10c}, this point of view displays remarkable
properties and many articles were devoted to this subject. In particular,
different authors studied the relations between the BFN approach and the
theory of deformation
quantization \cite{bfn11,bfn12,bfn13,bfn14,bfn15,bfn16}. N. C. Dias and
J. N. Prata study in their article \cite{bfn16} the relations between
quasi-distributions (the phase space representation of quantum states
in the deformation quantization approach) and the Bohm distributions in
phase space. As we will see in the appendix, our point of view allows us to
find a similar relation.

Now, let us situate our article. In a previous work
\cite{kvnberg} we have already explored to what extend
the KvN approach can be used to rebuild quantum mechanics using physical
arguments. The starting point of our study was not the abrupt data of the
KvN formalism, but rather a physical reasoning about the necessary
stochastic properties of classical observables. Namely we followed M. Born's
remark \cite{born} by noticing that the true mathematical representation of
any physical measurement is an expectation value associated with some
probability distribution, because a real result of a physical experiment
always contains some uncertainties. Therefore, physical continuous
observables are stochastic by nature. This important remark is also the
starting point of E. Prugove\v {c}ki's monograph \cite{coherentprugo1}.
Then the picture of classical mechanics based on a complete determinism is
an extrapolation of the real information obtained from experiments. But if
this picture cannot be proved from classical measurements, it cannot be
invalidated any more, in particular because classical equations are
compatible with (or based on) this picture. This explains the difficulty
to introduce, in a consistent way, intrinsic stochastic properties into the
classical framework. Nevertheless this stage is necessary, if we want to
find a continuous transition from the classical framework to the quantum one.
The KvN approach is one solution. It associates to each particle a square
integrable function  $\psi (\vec{p},\vec{q})$ on phase space such as the
probability distribution $\rho =\left| \psi (\vec{p},\vec{q})\right| ^{2}$
always exhibits uncertainties. We mean that the limit
$\rho \rightarrow \delta (\vec{p}-\vec{p}_{0},\vec{q}-\vec{q}_{0})$ is not
allowed, because there is no $\psi $ such as $\psi =\sqrt{\delta }$.
Then pure classical states ($\delta $ distributions) can be approximated,
but never reached. Then this formalism is a better representation of the real
information obtained from measurements.

In this article we want to explore another possibility to introduce
intrinsic stochastic properties into the classical framework, and also a
complete different form of continuous transition toward quantum mechanics.
Some parts of the article possess common technical features with the BFN
approach, but the leading ideas are very different, since we assume from the
beginning that physical observables are intrinsically stochastic at 
each given time. Then all interpretations in terms of trajectories (deduced
from our field equations) can only have a mathematical meaning.

One of the main idea is based on the analysis of the physical role of the
action minimization principle. This principle allows one to obtain dynamical
equations in a very large variety of situations. But this does not mean that
all physical equations of evolution are deductible from some action
minimization.

Particularly the equations of evolution of a pure state $\left( \vec{p}%
_{0}(t),\vec{q}_{0}(t)\right) $ can be deduced from an action minimization,
but we show in the section \ref{sec3.1} that the Liouville equation cannot
be obtained from a minimization principle. Nevertheless any solution of the
Liouville equation is a linear superposition of pure states, where the
coefficients are the probabilities (due to the observer). This suggests that
the action minimization principle only works for intrinsic dynamical
equations that do not include probabilities due to some observer. This leads
to the idea that the solutions of the Liouville equation, only parametrized
by quantities that verify some action minimization, are intrinsic dynamical
solutions, that is intrinsic stochastic distributions.

This allows us to define the $q$-stochastic pure states, intrinsic
stochastic distributions, analogous of the usual classical pure states.
These distributions are classical Bohm distributions usually 
interpreted as the classical limit of quantum pure states 
\cite{holland,bfn16}. They are parametrized by two fields, a probability
distribution $n(\vec{q},t)$ and an action field $S(\vec{q},t)$, such as
the equations verified by $n$ and $S$ are solutions of the variational
principle $\delta \int \mathcal{L}d^{3}\vec{q}dt=0$, where 
 $\mathcal{L}$ is the Lagrangian. [Of course these solutions can be classically
 interpreted in terms of trajectories as in the BFN approach, but in our
point of view these trajectories only possess a mathematical meaning.] The
properties of these states are analyzed in the section \ref{sec3.3}.

After the introduction of intrinsic stochastic properties into classical
mechanics, we analyze to what extend we can perform a transition to quantum
mechanics [Sec. \ref{sec4}]. This transition is obtained by a new
hypothesis, namely we must assume that the field $S$ exhibits intrinsic
stochastic properties. Then the classical Lagrangian $\mathcal{L}$ must be
replaced by some averaged Lagrangian $\mathcal{L}_{m}$.

We show in the section \ref{sec5} how $\mathcal{L}_{m}$ can be obtained, and
how it leads to the Schr\"{o}dinger equation. This derivation of the
Schr\"{o}dinger equation is technically close to the one already published
by M. J. W. Hall \cite{hallreginatto} and M. Reginatto \cite
{hallreginatto,reginatto1,reginatto2} (based on the Fisher information), but
our arguments are very different. This procedure also is different from the
ones proposed by M. Davidson \cite{davidson} or by G. Kaniadakis 
\cite{kaniadakis}.

Then we analyze in the section \ref{sec6} how all the quantum framework can
completely be rebuilt from the previous results. In particular, we show that
quantum axioms appear as abstract generalizations of classical calculations.

\section{CLASSICAL MECHANICS IN PHASE SPACE}

We recall in few lines the framework of classical mechanics \cite
{meca1,meca2}.

\subsection{Phase space and classical dynamics}

Phase space is the set of pairs $(\vec{p},\vec{q})$ of momenta and
positions, where $(\vec{p},\vec{q})$ represents the (pure) state of the
system. Physical observables are functions $f(\vec{p},\vec{q})$ on phase
space. If we call $M$ the $\mathbb{R}^{3}$ space manifold, phase space is
the cotangent bundle $TM^{*}$. It possesses a natural geometry (namely a
symplectic geometry) that allows us to define the Poisson brackets (PB), $%
\left\{ f,g\right\} $ of two functions by 
\begin{equation}
\left\{ f,g\right\} =\partial _{\vec{p}}f.\partial _{\vec{q}}g-\partial _{%
\vec{p}}g.\partial _{\vec{q}}f.  \label{PoissonBr}
\end{equation}

Dynamics on phase space is defined by the Hamiltonian equations 
\begin{subequations}
\label{hamilequs}
\begin{equation}
\frac{d}{dt}\vec{q}=\partial _{\vec{p}}H(\vec{q},\vec{p},t),
\end{equation}
\begin{equation}
\frac{d}{dt}\vec{p}=-\partial _{\vec{q}}H(\vec{q},\vec{p},t),
\end{equation}
\end{subequations}
where $H$ is the Hamiltonian (eventually time dependent).

\subsection{Dynamics and statistics}

These equations (\ref{hamilequs}) correspond to the ideal case of a particle
perfectly localized in phase space and we can represent this situation by
the probability distribution $\rho (\vec{p},\vec{q},t)=\delta \left( \vec{p}-%
\vec{p}_{0}(t)\right) \delta \left( \vec{q}-\vec{q}_{0}(t)\right) $. If we
build a general distribution $\rho $ as superposition of ``$\delta $'' by
defining $\rho =$ $\sum_{i}P_{i}\delta _{\vec{p}_{i}(t),\vec{q}_{i}(t)}$, we
find that $\rho $ verifies the Liouville equation: 
\begin{equation}
\frac{\partial \rho }{\partial t}=-\left\{ H,\rho \right\} \text{.}
\label{Liouville}
\end{equation}

We notice for the following that this equation can be written as 
\begin{equation}
\partial _{t}\rho +\partial _{\vec{q}}\left( \rho \partial _{\vec{p}
}H\right) -\partial _{\vec{p}}\left( \rho \partial _{\vec{q}}H\right) =0
\label{liouvillebis}
\end{equation}

So we say classically that Eq. (\ref{Liouville}) describes the evolution of
any probability distribution $\rho $.

Starting from a distribution $\rho $ that verifies the equation (\ref
{Liouville}), we can look at the evolution of the expectation value $%
\left\langle f\right\rangle _{t}$ of an observable $f(\vec{p},\vec{q},t)$
defined as 
\begin{equation}
\left\langle f\right\rangle _{t}=\int_{\mathbb{R}^{6}}f(\vec{p},\vec{q},t)\rho (%
\vec{q},\vec{p},t)d^{3}\vec{p}d^{3}\vec{q}.  \label{expectvalue}
\end{equation}
We find: 
\begin{equation}
\frac{d}{dt}\left\langle f\right\rangle _{t}=\left\langle \frac{\partial f}{%
\partial t}\right\rangle _{t}+\left\langle \left\{ H,f\right\} \right\rangle
_{t}.  \label{generalmeanequ}
\end{equation}

Applied to the special case of the two fundamental observables $\vec{p}$ and 
$\vec{q}$, Eq. (\ref{generalmeanequ}) gives: 
\begin{subequations} 
\label{meanhamilequs}
\begin{equation}
\frac{d}{dt}\left\langle \vec{q}\right\rangle _{t}=\left\langle \partial _{%
\vec{p}}H\right\rangle _{t}, 
\end{equation}
\begin{equation}
\frac{d}{dt}\left\langle \vec{p}\right\rangle _{t}=-\left\langle \partial _{%
\vec{q}}H\right\rangle _{t}.
\end{equation}  
\end{subequations}

\section{DEFINITION OF STOCHASTIC CLASSICAL PURE STATES\label{sec3}}

\subsection{Action minimization and the Liouville equation\label{sec3.1}}

As indicated above, classical pure states are defined as points $(\vec{p}%
_{0},\vec{q}_{0})$ of phase space corresponding to the ``$\delta $''
distribution $\rho _{0}(\vec{p},\vec{q})=\delta \left( \vec{p}-\vec{p}%
_{0}\right) \delta \left( \vec{q}-\vec{q}_{0}\right) $.\newline
We remark that the equations of evolution (\ref{hamilequs}) for such a pure
state (equivalent to the Liouville equation for $\rho _{0}$) are given by an
action minimization $\left( \delta \int Ldt=0\right) $ applied to the
Lagrangian $L$ depending on the parametric fields $\vec{q}_{0}(t)$ and $\vec{%
p}_{0}(t)$%
\begin{equation}
L\left( \vec{q}_{0},\frac{d\vec{q}_{0}}{dt},\vec{p}_{0},\frac{d\vec{p}_{0}}{%
dt},t\right) =\vec{p}_{0}.\frac{d\vec{q}_{0}}{dt}-H(\vec{q}_{0},\vec{p}%
_{0},t).
\end{equation}

In other words, usual pure states are special cases of parametrized
distributions $\rho $, solutions of the Liouville equation, such that the
dynamical equations verified by the parameters are the consequences of some
action minimization.

So we can wonder if other parametrized solutions of the Liouville equation
exhibiting the same property exist.

We must first verify that the Liouville equation itself cannot be obtained
from some action minimization [action based on a Lagrangian only depending
on the field $\rho $ and on the coordinates].\newline
To prove that, let us define a Lagrangian $\mathcal{L}(\rho ,\partial
_{\alpha }\rho ,\vec{p},\vec{q},t)$ [with $\alpha =\vec{p},$ $\vec{q}$ or $t$%
] associated with the minimization condition $\delta \int \mathcal{L}d^{3}%
\vec{p}d^{3}\vec{q}dt$ $=0.$ Then the equation in $\rho $ is given by 
\begin{equation}
\partial _{t}\frac{\partial \mathcal{L}}{\partial (\partial _{t}\rho )}%
+\partial _{\vec{p}}\frac{\partial \mathcal{L}}{\partial (\partial _{\vec{p}%
}\rho )}+\partial _{\vec{q}}\frac{\partial \mathcal{L}}{\partial (\partial _{%
\vec{q}}\rho )}=\frac{\partial \mathcal{L}}{\partial \rho }.  \label{minequa}
\end{equation}

If this equation (\ref{minequa}) reduces to the Liouville equation (\ref
{Liouville}), this implies that Eq. (\ref{minequa}) must not contain second
derivatives of $\rho $ in the variables $t$, $\vec{p}$ or $\vec{q}$. Then we
deduce 
\begin{subequations}
\begin{equation}
\frac{\partial \mathcal{L}}{\partial (\partial _{t}\rho )}=f(\rho ,\vec{p},%
\vec{q},t), 
\end{equation}
\begin{equation}
\frac{\partial \mathcal{L}}{\partial (\partial _{\vec{p}}\rho )}=\vec{g}%
(\rho ,\vec{p},\vec{q},t), 
\end{equation}
\begin{equation}
\frac{\partial \mathcal{L}}{\partial (\partial _{\vec{q}}\rho )}=\vec{h}%
(\rho ,\vec{p},\vec{q},t).
\end{equation}
\end{subequations}

So we have 
\begin{equation}
\mathcal{L}=F(\rho ,\vec{p},\vec{q},t)+f\partial _{t}\rho +\vec{g}.\partial
_{\vec{p}}\rho +\vec{h}.\partial _{\vec{q}}\rho .
\end{equation}

With this Lagrangian, the equation (\ref{minequa}) becomes 
\begin{equation}
\partial _{t}f(Y)+\partial _{p^{i}}g^{i}(Y)+\partial
_{q^{i}}h^{i}(Y)=\partial _{\rho }F(Y),
\end{equation}
where $Y=(\rho ,\vec{p},\vec{q},t)$.

Obviously this equation is not a differential equation in $\rho $, so the
Liouville equation cannot be obtained from some action minimization
based on a Lagrangian only depending on $\rho $ and the coordinates. [In
fact we can obtain the Liouville equation from a Lagrangian that contains
two unknown fields ($\rho $ and another one), but there is no reason to
introduce some supplementary field.]

\subsubsection{Conclusion\label{sec3.1.1}}

We have proved that only subsets of solutions of the Liouville equation are
the consequences of some action minimization. This result is important
because a general solution of the Liouville equation can be written as a
convex linear superposition of pure states where the coefficients are
interpreted as probabilities only due to the observer (the uncertainties are
not intrinsic).

The fact that pure states are derived from some action minimization, while a
general solution of the Liouville equation cannot be obtained by this
procedure seems to show that the action minimization principle discriminates
the solutions that are intrinsically dynamical from those that contain some
(or too many) external probabilities (due to the observer).

If we follow M. Born \cite{born} and E. Prugove\v {c}ki \cite{coherentprugo1}
by assuming that dynamical variables are intrinsically stochastic [exact
values must be seen as a mathematical extrapolation] we see that the action
minimization can be naturally lift up into a general procedure to specify
the solutions of the Liouville equation that are dynamical and intrinsically
stochastic (probabilities not due to the observer). We will call them \emph{%
\ stochastic pure states}. If these distributions exist, they constitute a
new starting point (similar to usual pure states) to rebuild a version of
classical mechanics that includes the idea of intrinsic stochastic
properties.

This is one of the main ideas of this article, and this explains why we are
looking for the solutions of the Liouville equation that verify this
condition.

\subsection{Solutions of the Liouville equation derived from an action
minimization\label{sec3.2}}

In the remainder we focus on two families of solutions (the proofs are given
below) defined as follows.

The first family (\textbf{I}) depends on the fields $n(\vec{q},t)$ and $S(%
\vec{q},t)$ and corresponds to classical Bohm distributions, usually 
interpreted as the classical limit of quantum pure states 
\cite{bfn16,holland}: 
\begin{equation}
\rho _{I}(\vec{p},\vec{q},t)=n(\vec{q},t)\delta \left( \vec{p}-\partial _{%
\vec{q}}S(\vec{q},t)\right) .  \label{parami}
\end{equation}

The second family (\textbf{II}) depends on the fields $n(\vec{p},t)$ and $S(%
\vec{p},t)$: 
\begin{equation}
\rho _{II}(\vec{p},\vec{q},t)=n(\vec{p},t)\delta \left( \vec{q}-\partial _{%
\vec{p}}S(\vec{p},t)\right) .  \label{paramii}
\end{equation}

\paragraph{Remark}

At first sight, these families can be unified and generalized, by
introducing some canonical transformation. But if we want to parametrize
distributions with fields that possess a direct physical meaning, these
fields must be defined on spaces that possess a material physical
realization (spaces where the Galilei group directly acts). Then, only the
first family (\textbf{I}) based on spacetime really is fundamental. The
second family (\textbf{II}) is its canonical conjugate. Other families
obtained after some canonical transformation that mixes $\vec{p}$ and $\vec{q%
}$ must be seen as purely mathematical solutions with no concrete physical
interest.

So the remainder of the article is essentially based on the analysis of the
solutions (\textbf{I}). Solutions (\textbf{II}) are analyzed at the end of
the article.

\subsubsection{Fields equations}

We begin first by the equations verified by the fields $n(\vec{q},t)$ and $S(%
\vec{q},t)$. If we write the Liouville equation (\ref{liouvillebis}) with
the formula (\ref{parami}), we obtain 
\begin{equation}
\partial _{t}n\delta -n\partial _{t\vec{q}}^{2}S.\partial _{\vec{p}}\delta
+\partial _{\vec{q}}\left( n\partial _{\vec{p}}H\delta \right) =\partial _{%
\vec{p}}\left( n\partial _{\vec{q}}H\delta \right) ,
\end{equation}
where $\delta =\delta (\vec{p}-\partial _{\vec{q}}S).$

Using the well-known identity verified by $\delta $ distributions $f(\vec{p},%
\vec{q})\delta (\vec{p}-\partial _{\vec{q}}S)=f(\partial _{\vec{q}}S,\vec{q}%
)\left. \delta (\vec{p}-\partial _{\vec{q}}S)\right. $, and after the
development of derivatives, we obtain 
\begin{eqnarray}
\partial _{t}n\delta -n\partial _{t\vec{q}}^{2}S.\partial _{\vec{p}}\delta
+\partial _{\vec{q}}\left( n\partial _{\vec{p}}H(\partial _{\vec{q}}S,\vec{q}%
,t))\right) \delta = \nonumber\\ 
n\partial p_{i}H(\partial _{\vec{q}}S,\vec{q},t)\partial
_{q_{i}q_{j}}^{2}S\partial _{p_{j}}\delta +n\partial _{\vec{q}}H(\partial _{%
\vec{q}}S,\vec{q},t).\partial _{\vec{p}}\delta ,
\end{eqnarray}
where the summation on $i$ and $j$ is implicit.

By collecting the terms in $\delta $ we obtain the equation for $n(\vec{q}
,t) $ that expresses the local conservation law

\begin{equation}
\partial _{t}n+\partial _{\vec{q}}\left( n\partial _{\vec{p}}H(\partial _{%
\vec{q}}S,\vec{q},t))\right) =0.  \label{nequai}
\end{equation}

By collecting the terms in $\partial _{p_{i}}\delta $, we obtain the
equation verified by $S(\vec{q},t)$

\begin{equation}
\partial _{\vec{q}}\left( \partial _{t}S+H(\partial _{\vec{q}}S,\vec{q}
,t)\right) =0.
\end{equation}

This equation can be simplified (up to a function of $t$), and we obtain the
Hamilton-Jacobi equation 
\begin{equation}
\partial _{t}S+H(\partial _{\vec{q}}S,\vec{q},t)=0.  \label{sequai}
\end{equation}

On the other hand we can verify that the equations (\ref{nequai}) and (\ref
{sequai}) are deduced from the variational principle $\delta \int \mathcal{L}
d^{3}\vec{q}dt=0$ where $\mathcal{L}$ is defined as 
\begin{equation}
\mathcal{L}=n\left( \partial _{t}S+H(\partial _{\vec{q}}S,\vec{q},t)\right) .
\label{lagrangian}
\end{equation}

\paragraph{Remark}

The definition of $\rho _{I}$ shows that the field $n$ is a probability
distribution on the ordinary space while $S$ is homogeneous to an action,
then the integral $\int \mathcal{L}d^{3}\vec{q}dt$ also is homogeneous to an
action. In addition, the classical Lagrangian $\mathcal{L}$ is often used
to describe a flow of classical particles, but in our approach it describes
the stochastic properties of a unique object, as we will see below.

\subsection{$q$-stochastic pure states\label{sec3.3}}

As previously mentioned, the distributions of type (\textbf{I}) can be 
interpreted as the classical limit of quantum pure states 
\cite{bfn16,holland}.
We also notice that the classical pure state
$\rho (\vec{p},\vec{q},t)=\left. \delta (\vec{p%
}-\vec{p}_{0}(t))\right. $ $\left. \delta (\vec{q}-\vec{q}_{0}(t))\right. $
correspond to the limit $n(\vec{q},t)\rightarrow \left. \delta (\vec{q}-\vec{%
q}_{0}(t))\right. $. Then pure states are limit cases of solutions of type (%
\textbf{I}). Moreover, if $\rho (\vec{p},\vec{q},t)$ is equal to $n(\vec{q}%
,t)\left. \delta (\vec{p}-\partial _{\vec{q}}S)\right. $, then for each
value of $\vec{q}$ only one value of $\vec{p}$ [equal to $\partial _{\vec{q}%
}S(\vec{q},t)$] is possible. So (as in the BFN approach) the function $S$
specifies a family of classical trajectories, solutions of the Hamiltonian
equations. These trajectories verify 
\begin{subequations}
\label{newtrajeceq}
\begin{equation}
\frac{d\vec{q}}{dt}=\partial _{\vec{p}}H(\partial _{\vec{q}}S,\vec{q},t), 
\end{equation}
\begin{equation}
\vec{p}=\partial _{\vec{q}}S(\vec{q},t).
\end{equation}
\end{subequations}
The function $n$ that represents the probability distribution on the $q$%
-space, also specifies the probability law of the trajectories (\ref
{newtrajeceq}) [since the data of $\vec{q}$ is sufficient to specify $\vec{p}
$]. But these trajectories only have a mathematical meaning in our approach,
since the position of the particle is assumed to be stochastic by nature.

So the distributions $\rho _{I}(\vec{p},\vec{q},t)=n(\vec{q},t)\left. \delta
(\vec{p}-\partial _{\vec{q}}S)\right. $ represent some extension of pure
states where only the condition of exact localization in the ordinary space
is relaxed, but a sharp localization is always possible. Moreover, as usual
pure states, the solutions (\textbf{I}) are the consequences of an action
minimization. We call the solutions of the family (\textbf{I}) $q$\emph{%
-stochastic pure states}.

\subsubsection{Conclusion}

Following the idea developed in the section \ref{sec3.1.1}, we assume in the
remainder that these $q$-stochastic pure states are intrinsic (classical)
stochastic states, stochastic analogous of pure states. So a particle must
be associated with a $q$-stochastic pure state at each given time.

On the other hand, convex linear superpositions of these $q$-stochastic pure
states always give a general probability distribution that verifies the
Liouville equation.

We deduce that classical statistical mechanics can be rebuilt starting from
the Lagrangian (\ref{lagrangian}) and the definition of the $q$-stochastic
pure states on phase space. But the usual formula for pure states must be
modified, and this leads to the following equations.

\subsection{Marginal laws and expectation values for $q$-stochastic pure
states\label{sec3.4}}

\subsubsection{The marginal laws}

For a probability distribution $\rho $ on phase space, the marginal laws
that gives the distributions $\mu (\vec{q},t)$ on the $q$-space and $\nu (%
\vec{p},t)$ on the $p$-space are given by 
\begin{equation}
\mu (\vec{q},t)=\int_{\mathbb{R}^{3}}\rho (\vec{p},\vec{q},t)d^{3}\vec{p}\text{%
; }\nu (\vec{p},t)=\int_{\mathbb{R}^{3}}\rho (\vec{p},\vec{q},t)d^{3}\vec{q}.
\end{equation}

For a $q$-stochastic pure state we obtain 
\begin{equation}
\mu (\vec{q},t)=n(\vec{q},t)\text{; }\nu (\vec{p},t)=\int_{\mathbb{R}%
^{3}}\delta \left( \vec{p}-\partial _{\vec{q}}S(\vec{q},t)\right) n(\vec{q}%
,t)d^{3}\vec{q}.  \label{marginalpq}
\end{equation}

\subsubsection{The expectation values}

For any observable $f(\vec{p},\vec{q})$ the expectation value $\left\langle
f\right\rangle _{t}$ is given by Eq. (\ref{expectvalue}), then 
\begin{equation}
\left\langle f\right\rangle _{t}=\int_{\mathbb{R}^{3}}f(\partial _{\vec{q}}S(%
\vec{q},t),\vec{q})n(\vec{q},t)d^{3}\vec{q}.
\end{equation}
The normalization condition becomes 
\begin{equation}
1=\left\langle 1\right\rangle _{t}=\int_{\mathbb{R}^{3}}n(\vec{q},t)d^{3}\vec{q}%
.  \label{normalization}
\end{equation}

The expectation values of position and momentum are 
\begin{subequations}
\label{expectpq}
\begin{equation}
\left\langle \vec{q}\right\rangle _{t}=\int_{\mathbb{R}^{3}}\vec{q}n(\vec{q}%
,t)d^{3}\vec{q}, 
\label{expectpqa}
\end{equation}
\begin{equation}
\left\langle \vec{p}\right\rangle _{t}=\int_{\mathbb{R}^{3}}\partial _{\vec{q}%
}S(\vec{q},t)n(\vec{q},t)d^{3}\vec{q}.
\label{expectpqb}
\end{equation}
\end{subequations}

The expectation values of energy and angular momentum are given by 
\begin{subequations}
\label{expecthl}
\begin{equation}
\left\langle H\right\rangle _{t}=\int_{\mathbb{R}^{3}}H(\partial _{\vec{q}}S(%
\vec{q},t),\vec{q})n(\vec{q},t)d^{3}\vec{q}, 
\label{expecthla}
\end{equation}
\begin{equation}
\left\langle \vec{l}\right\rangle =\left\langle \vec{q}\wedge \vec{p}%
\right\rangle _{t}=\int_{\mathbb{R}^{3}}\vec{q}\wedge \partial _{\vec{q}}S(\vec{%
q},t)n(\vec{q},t)d^{3}\vec{q}.
\label{expecthlb}
\end{equation}
\end{subequations}
The evolution of the expectation values of $\vec{p}$ and $\vec{q}$ follows
Eqs. (\ref{meanhamilequs}), then 
\begin{subequations}
\begin{equation}
\frac{d}{dt}\left\langle \vec{q}\right\rangle _{t}=\int_{\mathbb{R}^{3}}%
\partial _{\vec{p}}H(\partial _{\vec{q}}S,\vec{q},t)n(\vec{q},t)d^{3}\vec{q}, 
\end{equation}
\begin{equation}
\frac{d}{dt}\left\langle \vec{p}\right\rangle _{t}=-\int_{\mathbb{R}^{3}}%
\partial _{\vec{q}}H(\partial _{\vec{q}}S,\vec{q},t)n(\vec{q},t)d^{3}\vec{q}.
\end{equation}
\end{subequations}

\section{THE TRANSITION FROM CLASSICAL TO QUANTUM MECHANICS\label{sec4}}

\subsection{The hypothesis}

In the previous definition of $q$-stochastic pure states, stochastic
properties only appear through the probability distribution $n(\vec{q},t)$.
The stochastic properties of $\vec{p}$ are only due to those of $\vec{q}$,
because the well-defined function $S(\vec{q},t)$ specifies $\vec{p}$ for
each $\vec{q}$.

But if we assume that $S$ is an \emph{intrinsic} stochastic field (for each
given time), what are the consequences? This is the starting point of the
transition from classical to quantum mechanics.

The field $S$ can be essentially randomized starting from two different
hypothesis. The first hypothesis consists in assuming that $S$ is a random
field on the set of solutions of the Hamilton-Jacobi equation (equation
derived from the Lagrangian). The second hypothesis consists in assuming
that $S$ is a complete random field that does not follow any law.

The first hypothesis is not really consistent because the Hamilton-Jacobi
equation is derived from the Lagrangian $\mathcal{L}$ defined in Eq.(\ref
{lagrangian}). So, if we assume that $S$ is an intrinsic random field, then
the values of $\mathcal{L}$ also are randomized, and there is no reason to
assume that the Hamilton-Jacobi equation is always valid. So, only the
second hypothesis appears satisfactory, and we adopt this point of view in
the remainder.

\subsection{Consequences on the classical idea of trajectory}

We have seen that the $q$-stochastic pure state $\left( n,S\right) $ is
associated with a family of trajectories given by the equations (\ref
{newtrajeceq}). If we use the standard Hamiltonian $H=\frac{1}{2m}\vec{p}
^{2}+V(\vec{q})$, this leads to the equations 
\begin{subequations}
\label{trajecrandom}
\begin{equation}
\frac{d\vec{q}}{dt}=\frac{1}{m}\partial _{\vec{q}}S(\vec{q},t), 
\end{equation}
\begin{equation}
\vec{p}=\partial _{\vec{q}}S(\vec{q},t).
\end{equation}
\end{subequations}

If we assume that $S$ is a complete random field, and if we give a physical
meaning to the trajectories (this is not our case) then the previous
equations imply that all trajectories are possible (not restricted to the
Hamiltonian ones). Moreover for each given position $\vec{q}$, the momentum $%
\vec{p}$ becomes a complete random variable. So, on one hand we recover the
intuitive idea used by Feynmann to introduce the Feynmann path integral, and
on the other hand we have some kind of Heisenberg uncertainty principle.

\subsection{The quantization}

If $S(\vec{q},t)$ is a random field at each given time, then only the
expectation values of $S$ (or functions of $S$) have a physical meaning. So
the physical field becomes the averaged field $S_{m}(\vec{q},t)=\left\langle
S(\vec{q},t)\right\rangle $. The Lagrangian $\mathcal{L}=n\left( \partial
_{t}S+H(\partial _{\vec{q}}S,\vec{q},t)\right) $ is a physical quantity, so
it must be replaced by $\mathcal{L}_{m}=\left\langle \mathcal{L}%
\right\rangle $. It is natural to say that the averaged Lagrangian $\mathcal{%
L}_{m}$ only depends on the fields $n$ and $S_{m}$, because we assume that
the stochastic properties are intrinsic (they are not due to the action of
some unknown external dynamical field, or due to some observer effect).

Moreover, following our leading idea of section \ref{sec3.1.1}, if the
stochastic properties of $S$ are intrinsic, the action minimization
principle must always be valid, but now it must be applied to the Lagrangian 
$\mathcal{L}_{m}$ (then we assume implicitly that $\mathcal{L}_{m}$ depends
on the fields $n$ and $S_{m}$ through their values and their first
derivatives). The equations deduced from the new Lagrangian are the
quantized equations.

M. J. W. Hall and M. Reginatto \cite{hallreginatto,reginatto1,reginatto2}
obtained a derivation of the Schr\"{o}dinger equation from a procedure
technically close to this one.

\section{THE SCHRODINGER EQUATION\label{sec5}}

The principle of quantization previously developed depends explicitly on the
form of the Hamiltonian. So we assume in the remainder that $H$ is the
classical Hamiltonian 
\begin{equation}
H=\frac{1}{2m}\left( \vec{p}-e\vec{A}(\vec{q},t)\right) ^{2}+V(\vec{q},t).
\end{equation}
where $m$ is the mass and $e$ the charge of the particle.

The Lagrangian $\mathcal{L}$ is 
\begin{equation}
\mathcal{L}=n\left( \partial _{t}S+\frac{1}{2m}\left( \partial _{\vec{q}}S-e%
\vec{A}(\vec{q},t)\right) ^{2}+V(\vec{q},t)\right) .
\end{equation}

We notice that $\mathcal{L}$ is gauge invariant under the transformation 
\begin{equation}
S\rightarrow S+e\varphi ,\vec{A}\rightarrow \vec{A}+\partial _{\vec{q}
}\varphi ,V\rightarrow V-e\partial _{t}\varphi
\end{equation}

\subsection{The new Lagrangian $\mathcal{L}_{m}$}

The Lagrangian $\mathcal{L}_{m}=\left\langle \mathcal{L}\right\rangle $
introduced in the previous section becomes 
\begin{equation}
\mathcal{L}_{m}=n\left( \partial _{t}S_{m}+\frac{1}{2m}\left\langle \left(
\partial _{\vec{q}}S-e\vec{A}(\vec{q},t)\right) ^{2}\right\rangle +V(\vec{q}
,t)\right) ,
\end{equation}
where $S_{m}=\left\langle S\right\rangle $.

Moreover 
\begin{equation}
\left\langle \left( \partial _{\vec{q}}S-e\vec{A}(\vec{q},t)\right)
^{2}\right\rangle =\left( \partial _{\vec{q}}S_{m}-e\vec{A}(\vec{q}%
,t)\right) ^{2}+\Sigma ^{2},
\end{equation}
where $\Sigma =\sqrt{\left\langle \left( \partial _{\vec{q}}S-\partial _{%
\vec{q}}S_{m}\right) ^{2}\right\rangle }$ is the standard uncertainty.

We deduce that 
\begin{equation}
\mathcal{L}_{m}=\mathcal{L}(n,S_{m})+\frac{n}{2m}\Sigma ^{2},
\end{equation}
and the new Lagrangian $\mathcal{L}_{m}$ only is indeterminate through the
unknown uncertainty $\Sigma $.

\subsection{Obtaining $\Sigma $}

We list first the natural hypothesis that allows us to restrict the possible
forms of $\Sigma $.

\begin{description}
\item[(i)]  The dimensional analysis shows that $\Sigma $ is homogeneous to
an action divided by a length.

\item[(ii)]  If we assume that the Lagrangian $\mathcal{L}_{m}$ only depends
on $n$, $S_{m}$, their first derivatives and the spacetime coordinates, then 
$\Sigma $ has the same dependences. Since $\mathcal{L}$ only depends on the
coordinates through the external fields $\vec{A}$ and $V$, it is natural to
assume that this property is preserved in $\mathcal{L}_{m}$. But $\Sigma $
represents intrinsic stochastic effects (not due to external fields),
consequently it cannot depend on the external fields, and then it does not
depend on the coordinates. So $\Sigma $ is a function of $n$, $S_{m}$ and
their first derivatives.

\item[(iii)]  Since $\Sigma $ represents the effect of stochastic properties
of $S$ at each given time, it is logical to assume that $\Sigma $ does not
depend on time derivatives of $n$ and $S$.

\item[(iv)]  The Lagrangian $\mathcal{L}$ is invariant under gauge
transformations, and this symmetry is related to the conservation of 
particles,
so we assume that the gauge invariance is unbroken, and then $\Sigma $ must
be gauge invariant.
\end{description}

These simple remarks are sufficient to obtain the ``plausible'' form of $%
\Sigma $.

From (ii) and (iii), $\Sigma $ only depends on $\left( n,S_{m},\partial _{%
\vec{q}}n,\partial _{\vec{q}}S_{m}\right) $. But from (iv) we know that $%
\Sigma $ must be invariant under the transformation $S_{m}\rightarrow
S_{m}+e\varphi $. These two conditions are verified only if $\Sigma $ does
not depend on $S_{m}$. Then $\Sigma $ is a function of $n$ and $\partial _{%
\vec{q}}n$.

On the other hand, from (i) $\Sigma $ is homogeneous to an action divided by
a length, but it is impossible to build a quantity with this dimension only
using $n$ and $\partial _{\vec{q}}n$, because $n$ is homogeneous to $1/L^{3}$
and $\partial _{\vec{q}}n$ is homogeneous to $1/L^{4}$ (where $L$ is a
length). So we have to introduce some new constants.

Since we are looking for a quantity homogeneous to $a/L$ where ``$a$'' is an
action, the simplest idea is to introduce some unit of action ``$a$'' and
eventually some other undimensional (real) constant that we call ``$\beta $
''.

It is very easy to check that the simplest expression of $\Sigma ^{2}$
verifying these constraints is given by 
\begin{equation}
\Sigma ^{2}=a^{2}\left[ \left( \frac{\partial _{\vec{q}}n}{n}\right)
^{2}+\beta ^{2}n^{2/3}\right] .
\end{equation}
As we will see below the new constant ``$a$'' must be adjusted to $\hslash
/2 $, where $\hslash $ is the usual reduced Planck constant.

\subsubsection{Conclusion}

The uncertainty $\Sigma $ is given by the formula 
\begin{equation}
\Sigma ^{2}=\frac{\hslash ^{2}}{4}\left[ \left( \frac{\partial _{\vec{q}}n}{n%
}\right) ^{2}+\beta ^{2}n^{2/3}\right] ,
\end{equation}
where $\beta $ is some undimensional parameter.

We point out that this procedure gives a classical stochastic meaning to the
Planck constant $\hslash $ (as in the BFN approach), since this constant
only is related to some classical uncertainty. Moreover the limit $\hslash
\rightarrow 0$ means that the uncertainty $\Sigma $ vanishes and then $S$
cannot be any more a stochastic field.

On the other hand, we remark that our knowledge of the stochastic properties
of $S$ are limited to the following expectation values 
\begin{subequations}
\begin{equation}
\left\langle S(\vec{q},t)\right\rangle =S_{m}(\vec{q},t), 
\end{equation}
\begin{equation}
\left\langle (\partial _{\vec{q}}S)^{2}\right\rangle =(\partial _{\vec{q}
}S_{m})^{2}+\frac{1}{4}\hslash ^{2}\left[ \left( \frac{\partial _{\vec{q}}n}{
n}\right) ^{2}+\beta ^{2}n^{2/3}\right] .
\end{equation}
\end{subequations}

Nevertheless, from the knowledge of $\left\langle (\partial _{\vec{q}
}S)^{2}\right\rangle $, we can reasonably assume that the expectation value
of $\left. \partial _{q^{i}}S\partial _{q^{j}}S\right. $ verifies 
\begin{eqnarray}
\left\langle \partial _{q^{i}}S\partial _{q^{j}}S\right\rangle & = & \partial
_{q^{i}}S_{m}\partial _{q^{j}}S_{m} \nonumber\\
& & +\frac{1}{4}\hslash ^{2}\left[ \frac{
\partial _{q^{i}}n\partial _{q^{j}}n}{n^{2}}+\frac{1}{3}\delta _{ij}\beta
^{2}n^{2/3}\right] .  \label{meandsidsj}
\end{eqnarray}

We suppose that this condition is verified in the remainder.

As we will see below in the section \ref{sec6}, the partial knowledge of the 
$S$ stochastic properties selects the classical calculations that can be
performed, and this is a key point to show that we need some new axiom.

\subsection{The Schr\"{o}dinger equation}

We have the following Lagrangian $\mathcal{L}_{m}$%
\begin{equation}
\mathcal{L}_{m}=n\left( \partial _{t}S_{m}+\frac{1}{2m}\left( \partial _{%
\vec{q}}S_{m}-e\vec{A}\right) ^{2}+\frac{\Sigma ^{2}}{2m}+V\right) .
\end{equation}

We introduce the complex wave function $\psi (\vec{q},t)$ defined as 
\begin{equation}
\psi (\vec{q},t)=\sqrt{n}\exp \left( \frac{i}{\hslash }S_{m}\right) .
\label{wavefct}
\end{equation}

We find after some algebra 
\begin{eqnarray}
\mathcal{L}_{m}& = & \frac{i\hslash }{2}\left( \psi \partial _{t}\bar{\psi}-\bar{%
\psi}\partial _{t}\psi \right) +\frac{\hslash ^{2}}{2m}\overline{D_{\vec{q}%
}\psi }.D_{\vec{q}}\psi \nonumber\\
& & +\frac{\beta ^{2}\hslash ^{2}}{8m}\left| \psi
\right| ^{10/3}+V\left| \psi \right| ^{2},
\end{eqnarray}
where $\bar{\psi}$ represents the complex conjugate and $D_{\vec{q}}\psi
=\partial _{\vec{q}}\psi +i(e/\hslash )\vec{A}\psi $.

The minimization $\delta \int \mathcal{L}_{m}d^{3}\vec{q}dt=0$ can be done
on the fields $n$ and $S_{m}$, or equivalently on $\psi $ and $\bar{\psi}$.
This leads to the nonlinear Schr\"{o}dinger equation 
\begin{equation}
i\hslash \partial _{t}\psi =\frac{1}{2m}\left( -i\hslash \partial _{\vec{q}
}-e\vec{A}\right) ^{2}\psi +\frac{5\beta ^{2}\hslash ^{2}}{24m}\left| \psi
\right| ^{4/3}\psi +V\psi .
\end{equation}

If we assume that $\beta =0$, we recover the usual Schr\"{o}dinger equation.
In the remainder, we restrict ourselves to this special case.

\subsection{Conclusion}

This procedure allows us to recover the Schr\"{o}dinger equation from
classical stochastic arguments and the action minimization principle.
Moreover the semi-classical definition of the complex wave function appears
naturally, but its meaning is related to averaged fields.

In addition, $\psi $ is by construction the new version of the $q$%
-stochastic pure state. Then if a particle is represented by a $q$%
-stochastic pure state in classical mechanics, it must be represented by $%
\psi $ in the quantum theory [but the wave function is just a convenient
mathematical representation of the pair $\left( n,S_{m}\right) $].

We also notice that Bohm trajectories that are built from the action field 
$S_{m}$ can only be mathematical objects in our approach (since $S_{m}$ is an averaged field).

\section{THE QUANTUM FRAMEWORK\label{sec6}}

In this section we want to show how quantum ``axioms'' can be recovered in a
very natural way. From the previous section we already know that a particle
must be represented by a wave function that verifies the Schr\"{o}dinger
equation. Then it remains to study the relations linking the wave function,
the (classical) observables and the statistical techniques.

\subsection{Expectation values of the usual observables\label{sec6.1}}

The expectation values for $q$-stochastic pure states defined in the section 
\ref{sec3.4} were dependent on the fields $n$ and $S$. But if we assume that 
$S$ is a stochastic field, these expectation values must be also taken on $S$%
. This leads to the following equations for the usual observables.

The normalization condition [Eq. (\ref{normalization})] is unchanged 
\begin{equation}
1=\int_{\mathbb{R}^{3}}n(\vec{q},t)d^{3}\vec{q}=\int_{\mathbb{R}^{3}}\left| \psi
\right| ^{2}d^{3}\vec{q}.
\end{equation}

Then a particle must be represented by a normalized wave function.

The expectation value of position [Eq. (\ref{expectpqa})] gives 
\begin{equation}
\left\langle \vec{q}\right\rangle _{t}=\int_{\mathbb{R}^{3}}\vec{q}n(\vec{q}%
,t)d^{3}\vec{q}=\int_{\mathbb{R}^{3}}\vec{q}\left| \psi \right| ^{2}d^{3}\vec{q}%
.  \label{meanqQ}
\end{equation}

This expression can be generalized to any function of position 
\begin{equation}
\left\langle f(\vec{q})\right\rangle _{t}=\int_{\mathbb{R}^{3}}f(\vec{q})n(\vec{%
q},t)d^{3}\vec{q}=\int_{\mathbb{R}^{3}}f(\vec{q})\left| \psi \right| ^{2}d^{3}%
\vec{q}.  \label{meanfqQ}
\end{equation}

The expectation value of momentum [Eq. (\ref{expectpqb})] becomes (taking
into account vanishing integrals) 
\begin{equation}
\left\langle \vec{p}\right\rangle _{t}=\int_{\mathbb{R}^{3}}\partial _{\vec{q}%
}S_{m}n(\vec{q},t)d^{3}\vec{q}=-i\hslash \int_{\mathbb{R}^{3}}\bar{\psi}%
\partial _{\vec{q}}\psi d^{3}\vec{q}.  \label{meanpQ}
\end{equation}

We also find (taking into account Eq. (\ref{meandsidsj}) with the condition $%
\beta =0$) 
\begin{equation}
\left\langle p_{i}^{2}\right\rangle _{t}=\int_{\mathbb{R}^{3}}\left\langle (%
\partial _{q^{i}}S)^{2}\right\rangle n(\vec{q},t)d^{3}\vec{q}=\hslash
^{2}\int_{\mathbb{R}^{3}}\partial _{q^{i}}\bar{\psi}.\partial _{q^{i}}\psi d^{3}%
\vec{q}.  \label{meanppQ}
\end{equation}

The expectation value for the energy [Eq. (\ref{expecthla})] gives (with the
condition $\beta =0$) 
\begin{equation}
\left\langle H\right\rangle _{t}=\int_{\mathbb{R}^{3}}\left[ \frac{\hslash ^{2}%
}{2m}\overline{D_{\vec{q}}\psi }.D_{\vec{q}}\psi +V(\vec{q},t)\left| \psi
\right| ^{2}\right] d^{3}\vec{q}.  \label{meanhQ}
\end{equation}

Finally, the expectation value for angular momentum [Eq. (\ref{expecthlb} )]
becomes 
\begin{equation}
\left\langle \vec{l}\right\rangle _{t}=\int_{\mathbb{R}^{3}}\vec{q}\wedge 
\partial _{\vec{q}}S_{m}n(\vec{q},t)d^{3}\vec{q}=-i\hslash \int_{\mathbb{R}^{3}}%
\bar{\psi}\vec{q}\wedge \partial _{\vec{q}}\psi d^{3}\vec{q}.  \label{meanlQ}
\end{equation}

So we recover the usual quantum formula for the main classical observables
(that are in fact the generators of the Galilei group). But for the time
being, we have not introduced any quantum axiom about observables.

\paragraph{Remark}

The previous calculations do not prove the identity of classical and quantum
expectation values for any observable.

Because our approach only specifies the expectation values of $S$ and $%
(\partial _{\vec{q}}S)^{2}$, we cannot calculate expectation values with
classical formula for observables that are more than quadratic in $\vec{p}$.
This means, for example, that we cannot calculate the uncertainty on energy.
Facing this situation, two attitudes are possible:

\begin{description}
\item[(i)]  In a classical point of view, we must say that some important
properties of $S$ are unknown and then this theory is incomplete. So we have
to add some new information about $S$ by external means.

\item[(ii)]  In a quantum point of view, we must believe in the efficiency
of the action minimization principle and think that all the elements that we
have, are exactly that we need. The only way out is to assume a breakdown of
classical techniques. So we have to look for new consistent statistical
definitions, compatible with the formula that have been already found, that
allow us to calculate the classically undefined quantities. At first sight
this point of view can appear very formal, but it works.
\end{description}

Before going further in our approach, we open a parenthesis to study the
necessary conditions on the stochastic properties of $S$ that can be deduced
from the hypothesis of a complete simultaneous validity of quantum and
classical expectation value calculations.

\subsubsection{Compatibility conditions between classical and quantum
calculations\label{sec6.1.1}}

We have seen in Eq. (\ref{marginalpq}) the marginal laws $\mu (\vec{q},t)$
and $\nu (\vec{p},t)$ for a $q$-stochastic pure state. If we take into
account the stochastic properties of $S$, we obtain the marginal laws $\mu
_{C}(\vec{q},t)$ and $\nu _{C}(\vec{p},t)$ that must be used for
calculations with classical formula: 
\begin{eqnarray}
\mu _{C}(\vec{q},t) & = & n(\vec{q},t)\text{; } \nonumber\\
\nu _{C}(\vec{p},t) & = & \int_{\mathbb{R}%
^{3}}\left\langle \delta \left( \vec{p}-\partial _{\vec{q}}S(\vec{q}%
,t)\right) \right\rangle n(\vec{q},t)d^{3}\vec{q}.
\end{eqnarray}

On the other hand, for a given wave function $\psi $, we can find the
distribution laws $\mu _{Q}(\vec{q},t)$ and $\nu _{Q}(\vec{p},t)$ deduced
from quantum mechanics 
\begin{equation}
\mu _{Q}(\vec{q},t)=\left| \psi (\vec{q},t)\right| ^{2}\text{; }\nu _{Q}(%
\vec{p},t)=\left| \hat{\psi}(\vec{p},t)\right| ^{2},
\end{equation}
where $\hat{\psi}$ is the Fourier transform of $\psi $ defined as 
\begin{equation}
\hat{\psi}(\vec{p},t)=(2\pi \hslash )^{-3/2}\int_{\mathbb{R}^{3}}\exp
(-(i/\hslash )\vec{p}.\vec{q})\psi (\vec{q},t)d^{3}\vec{q}.
\end{equation}

If we take into account the definition of the wave function [Eq. (\ref
{wavefct})], and if we want the classical and quantum distributions to be
identical, we must have $\mu _{C}=\mu _{Q}$ and $\nu _{C}=\nu _{Q}$. The
condition $\mu _{C}=\mu _{Q}$ is verified since $\mu _{C}=\mu _{Q}=n$, then
it remains the compatibility condition $\nu _{C}=\nu _{Q}$ that gives 
\begin{equation}
\int_{\mathbb{R}^{3}}\left\langle \delta \left( \vec{p}-\partial _{\vec{q}%
}S\right) \right\rangle \left| \psi (\vec{q},t)\right| ^{2}d^{3}\vec{q}%
=\left| \hat{\psi}(\vec{p},t)\right| ^{2}.  \label{compatibil}
\end{equation}

This condition is a very strong statement about the stochastic properties of 
$S$. We analyze its consequences in the following lines.

First of all, we can take the Fourier transform of the previous equation,
and we obtain 
\begin{equation}
\int_{\mathbb{R}^{3}}\left\langle e^{-i\vec{\xi}.\partial _{\vec{q}%
}S}\right\rangle \left| \psi \right| ^{2}d^{3}\vec{q}=\int_{\mathbb{R}%
^{3}}e^{-i\vec{\xi}.\vec{p}}\left| \hat{\psi}\right| ^{2}d^{3}\vec{p},
\end{equation}
where the vector $\vec{\xi}$ is a mathematical parameter (independent of $%
\hslash $).

Then, the right-hand side of this equation can be transformed, leading to 
\begin{equation}
\int_{\mathbb{R}^{3}}e^{-i\vec{\xi}.\vec{p}}\left| \hat{\psi}\right| ^{2}d^{3}%
\vec{p}=\int_{\mathbb{R}^{3}}\bar{\psi}\left( \vec{q}+\frac{\hslash }{2}\vec{\xi%
}\right) \psi \left( \vec{q}-\frac{\hslash }{2}\vec{\xi}\right) d^{3}\vec{q}.
\end{equation}

We deduce that the expectation value $\left\langle \exp \left( -i\vec{\xi}.%
\partial _{\vec{q}}S\right) \right\rangle $ must verify 
\begin{eqnarray}
\int_{\mathbb{R}^{3}}& & \left\langle e^{-i\vec{\xi}.\partial _{\vec{q}%
}S}\right\rangle \left| \psi \right| ^{2} d^{3}\vec{q} = \nonumber\\
 \int_{\mathbb{R}^{3}} & & \bar{%
\psi}\left( \vec{q}+\frac{\hslash }{2}\vec{\xi}\right) \psi \left( \vec{q}-%
\frac{\hslash }{2}\vec{\xi}\right) d^{3}\vec{q}.  \label{compatibilend}
\end{eqnarray}

This last equation is the compatibility condition we are looking for.

At first sight, we can think that the solution is given by the relation 
\begin{equation}
\left\langle e^{-i\vec{\xi}.\partial _{\vec{q}}S}\right\rangle =\frac{1}{%
\left| \psi (\vec{q})\right| ^{2}}\bar{\psi}\left( \vec{q}+\frac{\hslash }{2}%
\vec{\xi}\right) \psi \left( \vec{q}-\frac{\hslash }{2}\vec{\xi}\right) ,
\end{equation}
but this expression is not satisfactory, although it possesses some right
properties. This particular point will be analyzed in the appendix, since it
does not interfere with the remainder of the article. In particular, 
we will show that this expression introduces a connection between Bohm 
distributions and Wigner functions. This point is in relation with 
the article \cite{bfn16} already mentioned.

We now return to our approach, and we analyze to what extend the previous
expressions of expectation values [Eqs. (\ref{meanqQ}) to (\ref{meanlQ})]
can imply the quantum axioms related to observables.

\subsection{The quantized observables}

We first introduce the Hilbert space $\mathcal{H}=\emph{L}^{2}(\mathbb{R}
^{3})$ equipped with the inner product $\left\langle \psi |\varphi
\right\rangle =\int_{\mathbb{R}^{3}}\bar{\psi}(\vec{q})\varphi (\vec{q})d^{3}%
\vec{q}$, and we define the self-adjoint operators $\vec{\bm{Q}}$ and%
\textbf{\ }$\vec{\bm{P}}$ as usually 
\begin{equation}
\vec{\bm{Q}}(\varphi )(\vec{q})=\vec{q}\varphi (\vec{q})\text{ and }%
\vec{\bm{P}}(\varphi )(\vec{q})=-i\hslash \partial _{\vec{q}}\varphi .
\end{equation}

\subsubsection{The quantum observables of position and momentum}

The expectation values of $\vec{p}$ and $\vec{q}$ given by the equations (%
\ref{meanpQ}) and (\ref{meanqQ}) can be written with the Dirac formalism as 
\begin{equation}
\left\langle \vec{p}\right\rangle =\left\langle \psi \left| \vec{\bm{P}}
\right| \psi \right\rangle \text{ and }\left\langle \vec{q}\right\rangle
=\left\langle \psi \left| \vec{\bm{Q}}\right| \psi \right\rangle .
\end{equation}

In a same way, from the equations (\ref{meanfqQ}) and (\ref{meanppQ}) we
obtain 
\begin{equation}
\left\langle p_{i}^{2}\right\rangle =\left\langle \psi \left| \bm{P}
_{i}^{2}\right| \psi \right\rangle \text{ and }\left\langle
q_{i}^{2}\right\rangle =\left\langle \psi \left| \bm{Q}_{i}^{2}\right|
\psi \right\rangle .
\end{equation}

Then we notice that not only the expectation values but also the
uncertainties for $\vec{p}$ and $\vec{q}$ correspond to the usual quantum
formula. This implies that the Heisenberg uncertainty principle is obtained
from classical stochastic arguments and without any quantum axiom. This
point is directly related to the articles of M. J. W. Hall and M. Reginatto 
\cite{hallreginatto,reginatto1,reginatto2} already mentioned. But this does
not mean that we do not need quantum axioms at all.

The equation (\ref{meanfqQ}) becomes 
\begin{equation}
\left\langle f(\vec{q})\right\rangle =\left\langle \psi \left| f(%
\vec{\bm{Q}})\right| \psi \right\rangle ,
\end{equation}
and this shows that the probability distribution of $\vec{q}$ is the
function $\left| \psi (\vec{q})\right| ^{2}$.

On the other hand, all the previous calculations do not directly specify the
probability distribution of the momentum $\vec{p}$ for the state $\psi $.
This means that we need external arguments to find this distribution.

If we introduce the eigenvectors $\left| \vec{p}\right\rangle $ of the
operator $\vec{\bm{P}}$ defined as 
\begin{equation}
\left| \vec{p}\right\rangle =(2\pi \hslash )^{-3/2}\int_{\mathbb{R}^{3}}\exp
[(i/\hslash )\vec{p}.\vec{q}]\left| \vec{q}\right\rangle d^{3}\vec{q},
\end{equation}
we find: 
\begin{equation}
\left\langle \vec{p}\right\rangle =\int_{\mathbb{R}^{3}}\vec{p}\left|
\left\langle \vec{p}|\psi \right\rangle \right| ^{2}d^{3}\vec{p}\text{ and }%
\left\langle p_{i}^{2}\right\rangle =\int_{\mathbb{R}^{3}}p_{i}^{2}\left|
\left\langle \vec{p}|\psi \right\rangle \right| ^{2}d^{3}\vec{p}.
\end{equation}

This means that the calculations of expectation values and uncertainties are
compatible with the interpretation of $\left| \left\langle \vec{p}|\psi
\right\rangle \right| ^{2}$ as being the probability distribution of
momentum.

\subsubsection{The angular momentum}

We define the self-adjoint operator $\vec{\bm{L}}$ as 
\begin{equation}
\vec{\bm{L}}=\vec{\bm{Q}}\wedge \vec{\bm{P}}.
\end{equation}

From Eq.(\ref{meanlQ}) we deduce 
\begin{equation}
\left\langle \vec{l}\right\rangle =\left\langle \vec{q}\wedge \vec{p}
\right\rangle =\left\langle \psi \left| \vec{\bm{L}}\right| \psi
\right\rangle .
\end{equation}

Taking into account the expectation value of $\left. \partial _{i}S\partial
_{j}S\right. $ [Eq. \ref{meandsidsj}] we also obtain 
\begin{equation}
\left\langle l_{i}^{2}\right\rangle =\left\langle \psi \left| \bm{L}
_{i}^{2}\right| \psi \right\rangle .
\end{equation}

Then the uncertainty $\Delta l_{i}$ on each component $l_{i}$ is given by
the quantum formula. Moreover we know by classical arguments that the
uncertainty $\Delta l_{i}$ vanishes when the expectation value $\left\langle
l_{i}\right\rangle $ is in fact an exact value. But the condition $\Delta
l_{i}=0$ means that $\psi $ is an eigenvector of $\bm{L}_{i}$ and that $%
\left\langle l_{i}\right\rangle $ is the correspondent eigenvalue. From
quantum algebraic calculations, we deduce that the possible values of the
angular momentum components are quantized. This result is obtained without
any quantum axiom. But this is not sufficient to specify the $l_{i}$
probability distribution.

Then we can develop a reasoning similar to the momentum case, and deduce
that the quantum formula of the probability distribution is compatible with
these results.

\subsubsection{The energy}

We define the self-adjoint operators $\bm{H}$ as 
\begin{equation}
\bm{H}=\frac{1}{2m}\left( \vec{\bm{P}}-e\vec{A}(\vec{\bm{Q}}
,t)\right) ^{2}+V(\vec{\bm{Q}},t).
\end{equation}

From Eq. (\ref{meanhQ}) we deduce 
\begin{equation}
\left\langle H\right\rangle =\left\langle \psi \left| \bm{H}\right| \psi
\right\rangle .
\end{equation}

As mentioned above in the section \ref{sec6.1}, our knowledge of the $S$
stochastic properties is not sufficient to specify the uncertainty for $H$,
but this quantity certainly exists. So we need to develop some new idea,
based on the results that we have already obtained.

\subsubsection{Axioms for the quantized observables\label{sec6.2.4}}

The previous calculations deal with the main physical observables, so it is
natural to try to generalize them axiomatically.

First, we can assume that any physical observable ``$a$'' is represented by
a self-adjoint operator ``$\bm{A}$ '' on the Hilbert space.

Second, the expectation value of ``$a$'' is given by $\left\langle
a\right\rangle =\left\langle \psi \left| \bm{A}\right| \psi
\right\rangle $.

These two axioms seem reasonable from the previous results, but they are not
sufficient to specify:\newline
- the probability distribution of the observable ``$a$'',\newline
- the operator $\bm{B}$ associated with the observable $f(a)$.

So we need some new arguments to reply to these questions. The momentum and
angular momentum studies give the method.

Since $\bm{A}$ is a self-adjoint operator, from the spectral theorem it
possesses the splitting 
\begin{equation}
\bm{A}=\int a\left| a,\alpha \right\rangle \left\langle a,\alpha \right|
\mu (a,\alpha )d\alpha da,
\end{equation}
where the kets $\left| a,\alpha \right\rangle $ are the eigenvectors
corresponding to the eigenvalue $a,$ and $\mu $ is some positive density
such that 
\begin{equation}
\int \left| a,\alpha \right\rangle \left\langle a,\alpha \right| \mu
(a,\alpha )d\alpha da=\bm{1}_{\mathcal{H}}.
\end{equation}

Then we can write 
\begin{subequations}
\label{expectgenQ}
\begin{equation}
\int \left\langle \psi |a,\alpha \right\rangle \left\langle a,\alpha |\psi
\right\rangle \mu (a,\alpha )d\alpha da=\left\langle \psi |\psi
\right\rangle =1, 
\end{equation}
\begin{equation}
\left\langle a\right\rangle =\left\langle \psi \left| \bm{A}\right| \psi
\right\rangle =\int a\left\langle \psi |a,\alpha \right\rangle \left\langle
a,\alpha |\psi \right\rangle \mu (a,\alpha )d\alpha da.
\end{equation}
\end{subequations}

If we define $\rho (a)=\int \left| \left\langle a,\alpha |\psi \right\rangle
\right| ^{2}\mu (a,\alpha )d\alpha $, the equations (\ref{expectgenQ}) show
that $\rho (a)$ can be classically interpreted as the probability
distribution of the observable ``$a$''. This constitutes the only new
ingredient that we need.

Then the expectation value of $f(a)$ is given by 
\begin{equation}
\left\langle f(a)\right\rangle =\int f(a)\rho (a)da=\left\langle \psi \left|
f(\bm{A})\right| \psi \right\rangle .
\end{equation}

We deduce that the operator associated with $f(a)$ must be $f(\bm{A})$.

With this new definition, we can calculate the uncertainty $\Delta a$ on ``$%
a $'' 
\begin{equation}
\Delta a^{2}=\left\langle a^{2}\right\rangle -\left\langle a\right\rangle
^{2}=\left\langle \psi \left| \bm{A}^{2}\right| \psi \right\rangle
-\left\langle \psi \left| \bm{A}\right| \psi \right\rangle ^{2}.
\end{equation}

Moreover, if $\rho _{\psi }(a)$ is the probability distribution of the
observable ``$a$'' for the state $\psi $, then the domain $Supp(\rho _{\psi
})$ of $\mathbb{R}$ where $\rho _{\psi }(a)$ does not vanish is by
definition the domain of possible values of ``$a$'' corresponding to the
state $\psi $. If we bring together the subsets $Supp(\rho _{\psi })$ for
all $\psi $, we must obtain the full set of possible values of ``$a$'' .
From the definition of $\rho _{\psi }$ we deduce that the set of possible
values is the spectrum of the operator $\bm{A}$.

\paragraph{Conclusion}

If we look at the concrete observables, this extended framework does not
change the expectation values and the uncertainties for the position and the
momentum (expressions obtained from classical formula), and these
observables retain a continuous spectrum. Only the probability distributions
of momentum have a new definition. Then we do not need quantum axioms to
recover the Heisenberg uncertainty principle.

We find without quantum axioms that the possible values of the angular
momentum are quantized, but we need one axiom to prove that the energy is
quantized (in suitable situations). Nevertheless the expectation values can
always be calculated using classical techniques.

So we remark that the quantum axiom adds new information (about
probability distributions) but do not invalidate the classical calculations
that can be done from our knowledge of the $S$ stochastic properties.

So if we have to specify what the quantum axiom about observables really
is, we must say that it is a new interpretation of classical expectation
value formula, leading to a new definition of the observable probability
distributions. This means that classical expectation value formula hide some
new possible statistical definition that can be taken into account, or
completely ignored. Nevertheless, this reasoning does not give the physical
meaning of this new interpretation.

We conclude this remark by noticing that the famous ``correspondence
principle'' is not needed in our approach.

In the list of usual quantum axioms, it remains two points that we need to
study. They concern the definition of a general probability distribution in
the quantum framework and the famous collapse of the wave function.

\subsection{Quantum densities and the collapse of the wave function}

\subsubsection{Quantum densities}

We mentioned in the section \ref{sec3.3} that a general classical
distribution $\rho $ on phase space is a convex linear superposition of $q$%
-stochastic pure states $\left\{ \nu _{\alpha }\right\} $%
\begin{equation}
\rho =\sum_{\alpha }p_{\alpha }\nu _{\alpha },
\end{equation}
with $\sum_{\alpha }p_{\alpha }=1$.

We deduce that the classical expectation values of $f(\vec{p},\vec{q})$
verify 
\begin{equation}
\left\langle f(\vec{p},\vec{q})\right\rangle _{\rho }=\sum_{\alpha
}p_{\alpha }\left\langle f(\vec{p},\vec{q})\right\rangle _{\nu _{\alpha }}.
\end{equation}

If each $q$-stochastic pure state $\nu _{\alpha }$ is represented in the new
framework by a wave function $\psi _{\alpha }$, and if the observable $f$
corresponds to the operator $\bm{F}$, we deduce that the expectation
value of $f$ (in the new framework) is given by 
\begin{equation}
\left\langle f\right\rangle =\sum_{\alpha }p_{\alpha }\left\langle \psi
_{\alpha }\left| \bm{F}\right| \psi _{\alpha }\right\rangle .
\end{equation}

Moreover if we define the operator $\bm{D}$ as 
\begin{equation}
\bm{D}=\sum_{\alpha }p_{\alpha }\left| \psi _{\alpha }\right\rangle
\left\langle \psi _{\alpha }\right| ,
\end{equation}
we have 
\begin{equation}
\left\langle f\right\rangle =Tr(\bm{D.F).}
\end{equation}

On the other hand, $\bm{D}$ is a positive operator and 
\begin{equation}
Tr(\bm{D})=\sum_{\alpha }p_{\alpha }=1.
\end{equation}

We deduce that a general statistical situation is represented by a positive
trace class operator $\bm{D}$ such that $Tr(\bm{D})=1$.

Reciprocally, any positive trace class operator $\bm{D}$ such that $Tr(%
\bm{D})=1$ describes a statistical situation, because it can be split in 
$\bm{D}=\sum_{n}p_{n}\left| \psi _{n}\right\rangle \left\langle \psi
_{n}\right| $ where $p_{n}\geq 0$ and $\sum_{n}p_{n}=1$.

\subsubsection{The collapse of the wave function}

In the spirit of our approach, the collapse of the wave function is just a
quantum translation of conditional probabilities. So let us briefly recall
what conditional probabilities in the classical framework are.

If we have a random variable ``$a$'' described by the probability
distribution $\rho (a)$ and if we do some experiment $\mathcal{E}$ that
specifies that the only possible values of ``$a$'' belong to some subset $%
\Delta $ of $\mathbb{R}$, the new probability distribution $\rho _{\mathcal{E%
}}(a)$ after the experiment is given by the conditional probability law 
\begin{equation}
\rho _{\mathcal{E}}(a)=\frac{\chi _{\Delta }(a)\rho (a)}{\int_{\Delta }\rho
(x)dx},  \label{conditionproba}
\end{equation}
where $\chi _{\Delta }$ is defined by $\chi _{\Delta }(x)=1$ if $x\in \Delta 
$ and $\chi _{\Delta }(x)=0$ elsewhere.

We assume now that ``$a$'' is a physical observable represented by the
self-adjoint operator $\bm{A}$ and we look at a particle in the (pure)
state $\psi $. The reasoning of the section \ref{sec6.2.4} shows that the
probability distribution $\rho (a)$ (in the usual sense) that describes the
stochastic properties of ``$a$'' is given by 
\begin{equation}
\rho (a)=\int \left| \left\langle a,\alpha |\psi \right\rangle \right|
^{2}\mu (a,\alpha )d\alpha ,
\end{equation}
where the notations are those of the section \ref{sec6.2.4}.

If we do some experiment $\mathcal{E}$ that specifies that the range of
values of ``$a$'' is some subset $\Delta $ of $\mathbb{R}$, we deduce that
after the experiment the probability law is the conditional probability $%
\rho _{\mathcal{E}}(a)$ given by the equation (\ref{conditionproba}).

But after the experiment the particle must always be described by some
normalized wave function $\psi _{\mathcal{E}}$ since the particle is
associated with a wave function for each given time (the experiment is
assumed to be perfect). Then $\psi _{\mathcal{E}}$ must verify 
\begin{equation}
\rho _{\mathcal{E}}(a)=\int \left| \left\langle a,\alpha |\psi _{\mathcal{E}%
}\right\rangle \right| ^{2}\mu (a,\alpha )d\alpha ,
\end{equation}
with 
\begin{equation}
\rho _{\mathcal{E}}(a)=\frac{\chi _{\Delta }(a)}{\int_{\Delta }\rho (x)dx}%
\int \left| \left\langle a,\alpha |\psi \right\rangle \right| ^{2}\mu
(a,\alpha )d\alpha .
\end{equation}

This condition is fulfilled for $\psi _{\mathcal{E}}$ defined as 
\begin{equation}
\left| \psi _{\mathcal{E}}\right\rangle =\frac{1}{\sqrt{\left\langle \psi
\left| \Pi _{\Delta }\right| \psi \right\rangle }}\Pi _{\Delta }\left| \psi
\right\rangle ,
\end{equation}
where $\Pi _{\Delta }$ is the projector 
\begin{equation}
\Pi _{\Delta }=\int_{a\in \Delta }\left| a,\alpha \right\rangle \left\langle
a,\alpha \right| \mu (a,\alpha )d\alpha da.
\end{equation}

Then we recover the collapse of the wave function.

So in our approach, the collapse of the wave function appears as a
mathematical tool that translates the (classical) conditional probability
process into quantum language. But it is well-known that the logical
consequences in the quantum framework are very different from those in the
classical framework.

We end this article by a last section devoted to the canonical conjugate
approach mentioned in Sec. \ref{sec3.2}.

\section{THE CANONICAL CONJUGATE APPROACH}

As indicated in the section \ref{sec3.2}, we can start the reasoning from
another family of stochastic pure states that we call $p$-stochastic pure
states defined as 
\begin{equation}
\rho (\vec{p},\vec{q},t)=n(\vec{p},t)\delta \left( \vec{q}-\partial _{\vec{p}%
}S(\vec{p},t)\right) .
\end{equation}

We briefly summarize the different steps of the study, similar to those
already developed.

Assuming that $\rho $ verifies the Liouville equation, we obtain the
equations of evolution for $n$ and $S$%
\begin{subequations}
\begin{equation}
\partial _{t}n-\partial _{\vec{p}}\left( n\partial _{\vec{q}}H(\vec{p}
,\partial _{\vec{p}}S)\right) =0, 
\end{equation}
\begin{equation}
\partial _{t}S-H(\vec{p},\partial _{\vec{p}}S)=0.
\end{equation}
\end{subequations}

These equations are solutions of the minimization condition $\delta \int 
\mathcal{L}d^{3}\vec{p}dt=0$, with $\mathcal{L}$ defined as 
\begin{equation}
\mathcal{L}=n\left( \partial _{t}S-H(\vec{p},\partial _{\vec{p}}S)\right) .
\end{equation}

Each $p$-stochastic pure state is associated with a family of solutions of
the Hamiltonian equations verifying 
\begin{subequations}
\begin{equation}
\vec{q}=\partial _{\vec{p}}S(\vec{p},t), 
\end{equation}
\begin{equation}
\frac{d\vec{p}}{dt}=-\partial _{\vec{q}}H(\vec{p},\partial _{\vec{p}}S).
\end{equation}
\end{subequations}

The field $n(\vec{p},t)$ is a probability distribution on the $p$-space, and
it also specifies the probability law of these trajectories.

The normalization condition becomes 
\begin{equation}
1=\left\langle 1\right\rangle _{t}=\int_{\mathbb{R}^{3}}n(\vec{p},t)d^{3}\vec{p}%
.
\end{equation}

The expectation values of position and momentum are given by 
\begin{subequations}
\begin{equation}
\left\langle \vec{q}\right\rangle _{t}=\int_{\mathbb{R}^{3}}n(\vec{p},t)%
\partial _{\vec{p}}Sd^{3}\vec{p}, 
\end{equation}
\begin{equation}
\left\langle \vec{p}\right\rangle _{t}=\int_{\mathbb{R}^{3}}\vec{p}n(\vec{p}%
,t)d^{3}\vec{p},
\end{equation}
\end{subequations}
moreover the expectation values of energy and angular momentum are 
\begin{subequations}
\begin{equation}
\left\langle H\right\rangle _{t}=\int_{\mathbb{R}^{3}}H(\vec{p},\partial _{\vec{%
p}}S)n(\vec{p},t)d^{3}\vec{p}, 
\end{equation}
\begin{equation}
\left\langle \vec{q}\wedge \vec{p}\right\rangle _{t}=\int_{\mathbb{R}^{3}}%
\partial _{\vec{p}}S\wedge \vec{p}n(\vec{p},t)d^{3}\vec{p}.
\end{equation}
\end{subequations}

Then we start the quantization process by assuming that $S$ is a stochastic
field, and we define the expectation value $S_{m}(\vec{p},t)=\left\langle S(%
\vec{p},t)\right\rangle $. The following step is the calculation of the
Lagrangian expectation value $\mathcal{L}_{m}$, where the Hamiltonian $H$ is
always the classical one. This leads to the equation 
\begin{equation}
\mathcal{L}_{m} =n\left( \frac{1}{2m}
\left\langle \left( \vec{p}-e\vec{A}(\partial _{\vec{p}}S,t)\right)
^{2}\right\rangle +\left\langle V(\partial _{\vec{p}}S,t)\right\rangle
\right) .
\end{equation}

But unlike the $q$-stochastic case, we generally need a complete knowledge
of the stochastic properties of $S$ to calculate this expectation value. So
the procedure can only be continued in very special situations.

The conclusion is that the roles of $\vec{p}$ and $\vec{q}$ cannot be
reversed in our reasoning, because the symmetry in $\left( \vec{p},\vec{q}
\right) $ of the classical mechanics formulation in phase space only is a
mathematical appearance. This symmetry does not take into account the fact
that the classical Hamiltonian is not any function of $\vec{p}$ and $\vec{q}$
, but a specific one that reflects the physical meaning of the dynamical
variables.

\section{CONCLUSION}

All the elements of the quantum framework are rebuilt, only starting from
the classical framework and some stochastic ingredients. Physical arguments
and natural generalizations of classical formula only are needed.

This shows that the usual axiomatic approach to quantum mechanics can
partially be bypassed, allowing us to obtain a picture of quantum mechanics
closer to classical (statistical) mechanics, even if the quantum equations
are finally the usual ones (but we notice that the linear Schr\"{o}dinger
equation is not the unique possibility). Moreover, different key elements of
the quantum formalism (the Heisenberg uncertainty principle, the
correspondence principle, the quantization of angular momentum) are obtained
from classical stochastic reasonings.

Nevertheless we remark that one axiom (about probability distribution of
observables) always is necessary. If the necessity of this axiom clearly
appears on a logical level, its physical interpretation from 
classical framework remains obscure.

\appendix 
\section{}
This appendix is devoted to the study of the compatibility condition
introduced in the section \ref{sec6.1.1}. We recall that the stochastic
field $S$ must verify the equation (\ref{compatibilend}), and a possible
solution is given by 
\begin{equation}
\left\langle e^{-i\vec{\xi}.\partial _{\vec{q}}S}\right\rangle =\frac{1}{%
\left| \psi (\vec{q})\right| ^{2}}\bar{\psi}\left( \vec{q}+\frac{\hslash }{2}%
\vec{\xi}\right) \psi \left( \vec{q}-\frac{\hslash }{2}\vec{\xi}\right) . 
\end{equation}

This expression possesses the right symmetry for the complex conjugation,
and we want to see to what extend it is relevant. Then we study the
development of the expectation value in power of $\vec{\xi}$ to the second
order. We have 
\begin{equation}
\left\langle e^{-i\vec{\xi}.\partial _{\vec{q}}S}\right\rangle =1-i\xi
^{i}\left\langle \partial _{i}S\right\rangle -\frac{1}{2}\xi ^{i}\xi
^{j}\left\langle \partial _{i}S\partial _{j}S\right\rangle +o(\xi ^{2}). 
\label{dlexpect}
\end{equation}

Then, we can do the same development for $\psi $%
\begin{eqnarray}
\psi \left( \vec{q}-\frac{\hslash }{2}\vec{\xi}\right) =& & \psi (\vec{q})-\frac{
\hslash }{2}\xi ^{i}\partial _{i}\psi (\vec{q}) \nonumber\\
& & +\frac{\hslash ^{2}}{8}\xi
^{i}\xi ^{j}\partial _{ij}^{2}\psi (\vec{q})+o(\xi ^{2}). 
\end{eqnarray}

We deduce 
\begin{eqnarray}
\frac{\bar{\psi}\left( \vec{q}+\frac{1}{2}\hslash \vec{\xi}\right) \psi
\left( \vec{q}-\frac{1}{2}\hslash \vec{\xi}\right) }{\left| \psi (\vec{q}%
)\right| ^{2}}= 1+\xi ^{i}M_{i} \nonumber\\
+\frac{1}{2}\xi ^{i}\xi ^{j}M_{ij}+o(\xi ^{2}), \label{dlexpectb}
\end{eqnarray}
with 
\begin{subequations}
\begin{equation}
M_{i}=\frac{1}{2\left| \psi (\vec{q})\right| ^{2}}\hslash \left( \psi
\partial _{i}\bar{\psi}-\bar{\psi}\partial _{i}\psi \right) , 
\end{equation}
\begin{equation}
M_{ij}= \frac{1}{8\left| \psi (\vec{q})\right| ^{2}}\hslash ^{2}\left( \bar{%
\psi}\partial _{ij}^{2}\psi +\psi \partial _{ij}^{2}\bar{\psi}-\partial _{i}%
\bar{\psi}\partial _{j}\psi -\partial _{i}\psi \partial _{j}\bar{\psi}%
\right) .
\end{equation}
\end{subequations}

By identification of the formula (\ref{dlexpect}) and (\ref{dlexpectb}) we obtain 
\begin{subequations}
\label{expectSb}
\begin{equation}
\left\langle \partial _{i}S\right\rangle =\frac{i}{2\left| \psi (\vec{q}%
)\right| ^{2}}\hslash \left( \psi \partial _{i}\bar{\psi}-\bar{\psi}\partial
_{i}\psi \right) , 
\end{equation}
\begin{eqnarray}
\left\langle \partial _{i}S\partial _{j}S\right\rangle = & & \left( \partial _{i}\bar{\psi}%
\partial _{j}\psi+\partial _{i}\psi \partial _{j}\bar{\psi}-\bar{\psi}%
\partial _{ij}^{2}\psi -\psi \partial _{ij}^{2}\bar{\psi}\right) \nonumber\\
& & \times \frac{1}{4\left|\psi (\vec{q})\right| ^{2}}\hslash ^{2}.
\end{eqnarray}
\end{subequations}

If we take into account the definition of the wave function as 
\begin{equation}
\psi =\sqrt{n}\exp \left( \frac{i}{\hslash }S_{m}\right) ,
\end{equation}
the expectation values of the equations (\ref{expectSb}) become 
\begin{subequations}
\begin{equation}
\left\langle \partial _{i}S\right\rangle =\partial _{i}S_{m}, 
\end{equation}
\begin{equation}
\left\langle \partial _{i}S\partial _{j}S\right\rangle =\partial
_{i}S_{m}\partial _{j}S_{m}-\frac{1}{4}\hslash ^{2}\partial _{ij}^{2}\ln (n).
\end{equation}
\end{subequations}

Then we recover the expectation value of $S$, but we find 
\begin{equation}
\left\langle \partial _{i}S\partial _{j}S\right\rangle =\partial
_{i}S_{m}\partial _{j}S_{m}+\frac{1}{4}\hslash ^{2}\frac{\partial
_{i}n\partial _{j}n}{n^{2}}-\frac{1}{4}\hslash ^{2}\frac{\partial _{ij}^{2}n%
}{n}.
\end{equation}

So we do not recover the right expectation value, due to the term in $%
\partial _{ij}^{2}n$. Moreover the standard uncertainty is not always a
positive number. But the supplementary term disappears in the integral $\int
n\left\langle (\partial _{\vec{q}}S)^{2}\right\rangle d^{3}\vec{q}$, then it
does not contribute to the integral form of the compatibility condition
specified by the equation (\ref{compatibilend}).

We end this appendix by noticing the relation that exists between the
proposed solution and the Wigner functions. Namely the expression of $%
\left\langle \exp \left( -i\vec{\xi}.\partial _{\vec{q}}S\right)
\right\rangle $ implies that $n(\vec{q})\left\langle \delta (\vec{p}-%
\partial _{\vec{q}}S)\right\rangle $ is equal to 
\begin{equation}
\int_{\mathbb{R}^{3}}\exp [(i/\hslash )\vec{p}.\vec{x}]\bar{\psi}(\vec{q}+\vec{x%
}/2)\psi (\vec{q}-\vec{x}/2)\frac{d^{3}\vec{x}}{(2\pi \hslash )^{3}}. 
\end{equation}

So we find that $n(\vec{q})\left\langle \delta (\vec{p}-\partial _{\vec{q}%
}S)\right\rangle $ is the Wigner function associated with the projector $%
\left| \psi ><\psi \right| $. But it is well-known that this Wigner function
is not always positive, then it cannot be equal to a probability
distribution. Nevertheless this shows the existence of some relation between
the averaged classical Bohm distributions and the Wigner functions 
(quasi-distributions), and this is directly related to the article of N. 
C. Dias and J. N. Plata \cite{bfn16}.

The conclusion of this appendix is that the local solution proposed for the
compatibility condition is not satisfactory, even if it is not so very far
from the solution we are looking for.
\bibliography{deriv}

\begin{thebibliography}{53}
\expandafter\ifx\csname natexlab\endcsname\relax\def\natexlab#1{#1}\fi
\expandafter\ifx\csname bibnamefont\endcsname\relax
  \def\bibnamefont#1{#1}\fi
\expandafter\ifx\csname bibfnamefont\endcsname\relax
  \def\bibfnamefont#1{#1}\fi
\expandafter\ifx\csname citenamefont\endcsname\relax
  \def\citenamefont#1{#1}\fi
\expandafter\ifx\csname url\endcsname\relax
  \def\url#1{\texttt{#1}}\fi
\expandafter\ifx\csname urlprefix\endcsname\relax\def\urlprefix{URL }\fi
\providecommand{\bibinfo}[2]{#2}
\providecommand{\eprint}[2][]{\url{#2}}

\bibitem[{\citenamefont{Weyl}(1927)}]{wignerweyl1}
\bibinfo{author}{\bibfnamefont{H.}~\bibnamefont{Weyl}}, \bibinfo{journal}{Z.
  Phys.} \textbf{\bibinfo{volume}{40}}, \bibinfo{pages}{1}
  (\bibinfo{year}{1927}).

\bibitem[{\citenamefont{Wigner}(1932)}]{wignerweyl2}
\bibinfo{author}{\bibfnamefont{E.}~\bibnamefont{Wigner}},
  \bibinfo{journal}{Phys. Rev.} \textbf{\bibinfo{volume}{40}},
  \bibinfo{pages}{749} (\bibinfo{year}{1932}).

\bibitem[{\citenamefont{Baker}(1958)}]{wignerweyl3}
\bibinfo{author}{\bibfnamefont{G.~A.} \bibnamefont{Baker}},
  \bibinfo{journal}{Phys. Rev.} \textbf{\bibinfo{volume}{109}},
  \bibinfo{pages}{2196} (\bibinfo{year}{1958}).

\bibitem[{\citenamefont{Moyal}(1949)}]{wignerweyl4}
\bibinfo{author}{\bibfnamefont{J.~E.} \bibnamefont{Moyal}},
  \bibinfo{journal}{Proc. Cambridge Philos. Soc.}
  \textbf{\bibinfo{volume}{45}}, \bibinfo{pages}{99} (\bibinfo{year}{1949}).

\bibitem[{\citenamefont{Balazs and Jennings}(1984)}]{wignerweyl5}
\bibinfo{author}{\bibfnamefont{N.~L.} \bibnamefont{Balazs}} \bibnamefont{and}
  \bibinfo{author}{\bibfnamefont{B.~K.} \bibnamefont{Jennings}},
  \bibinfo{journal}{Phys. Rep.} \textbf{\bibinfo{volume}{104}},
  \bibinfo{pages}{347} (\bibinfo{year}{1984}).

\bibitem[{\citenamefont{Hillery et~al.}(1984)\citenamefont{Hillery, O'Connel,
  Scully, and Wigner}}]{wignerweyl6}
\bibinfo{author}{\bibfnamefont{M.}~\bibnamefont{Hillery}},
  \bibinfo{author}{\bibfnamefont{R.~F.} \bibnamefont{O'Connel}},
  \bibinfo{author}{\bibfnamefont{M.}~\bibnamefont{Scully}}, \bibnamefont{and}
  \bibinfo{author}{\bibfnamefont{E.}~\bibnamefont{Wigner}},
  \bibinfo{journal}{Phys. Rep.} \textbf{\bibinfo{volume}{106}},
  \bibinfo{pages}{121} (\bibinfo{year}{1984}).

\bibitem[{\citenamefont{Bayen et~al.}(1978{\natexlab{a}})\citenamefont{Bayen,
  Flato, Fronsdal, Lichnerowicz, and Sternheimer}}]{starproduct1}
\bibinfo{author}{\bibfnamefont{F.}~\bibnamefont{Bayen}},
  \bibinfo{author}{\bibfnamefont{M.}~\bibnamefont{Flato}},
  \bibinfo{author}{\bibfnamefont{C.}~\bibnamefont{Fronsdal}},
  \bibinfo{author}{\bibfnamefont{A.}~\bibnamefont{Lichnerowicz}},
  \bibnamefont{and}
  \bibinfo{author}{\bibfnamefont{D.}~\bibnamefont{Sternheimer}},
  \bibinfo{journal}{Ann. Phys. (Paris)} \textbf{\bibinfo{volume}{110}},
  \bibinfo{pages}{111} (\bibinfo{year}{1978}{\natexlab{a}}).

\bibitem[{\citenamefont{Bayen et~al.}(1978{\natexlab{b}})\citenamefont{Bayen,
  Flato, Fronsdal, Lichnerowicz, and Sternheimer}}]{starproduct2}
\bibinfo{author}{\bibfnamefont{F.}~\bibnamefont{Bayen}},
  \bibinfo{author}{\bibfnamefont{M.}~\bibnamefont{Flato}},
  \bibinfo{author}{\bibfnamefont{C.}~\bibnamefont{Fronsdal}},
  \bibinfo{author}{\bibfnamefont{A.}~\bibnamefont{Lichnerowicz}},
  \bibnamefont{and}
  \bibinfo{author}{\bibfnamefont{D.}~\bibnamefont{Sternheimer}},
  \bibinfo{journal}{Ann. Phys. (Paris)} \textbf{\bibinfo{volume}{111}},
  \bibinfo{pages}{61} (\bibinfo{year}{1978}{\natexlab{b}}).

\bibitem[{\citenamefont{Flato et~al.}(1975)\citenamefont{Flato, Lichnerowicz,
  and Sternheimer}}]{starproduct3}
\bibinfo{author}{\bibfnamefont{M.}~\bibnamefont{Flato}},
  \bibinfo{author}{\bibfnamefont{A.}~\bibnamefont{Lichnerowicz}},
  \bibnamefont{and}
  \bibinfo{author}{\bibfnamefont{D.}~\bibnamefont{Sternheimer}},
  \bibinfo{journal}{Composito Mathematica} \textbf{\bibinfo{volume}{31}},
  \bibinfo{pages}{41} (\bibinfo{year}{1975}).

\bibitem[{\citenamefont{Perelomov}(1972)}]{coherentstates1}
\bibinfo{author}{\bibfnamefont{A.~M.} \bibnamefont{Perelomov}},
  \bibinfo{journal}{Commun. Math.} \textbf{\bibinfo{volume}{26}},
  \bibinfo{pages}{222} (\bibinfo{year}{1972}).

\bibitem[{\citenamefont{Klauder}(1979)}]{coherentstates2}
\bibinfo{author}{\bibfnamefont{J.~R.} \bibnamefont{Klauder}},
  \bibinfo{journal}{Phys. Rev. D} \textbf{\bibinfo{volume}{19}},
  \bibinfo{pages}{2349} (\bibinfo{year}{1979}).

\bibitem[{\citenamefont{Klauder}(1985)}]{coherentstates3}
\bibinfo{author}{\bibfnamefont{J.~R.} \bibnamefont{Klauder}},
  \emph{\bibinfo{title}{Coherent States. Applications in Physics and
  Mathematical Physics}} (\bibinfo{publisher}{World Scientific},
  \bibinfo{address}{Singapore}, \bibinfo{year}{1985}).

\bibitem[{\citenamefont{Perelomov}(1986)}]{coherentstates4}
\bibinfo{author}{\bibfnamefont{A.~M.} \bibnamefont{Perelomov}},
  \emph{\bibinfo{title}{Generalized Coherent States and Their Applications}}
  (\bibinfo{publisher}{Springer}, \bibinfo{address}{Berlin},
  \bibinfo{year}{1986}).

\bibitem[{\citenamefont{Bergeron and Valance}(1995)}]{coherentstatesberg}
\bibinfo{author}{\bibfnamefont{H.}~\bibnamefont{Bergeron}} \bibnamefont{and}
  \bibinfo{author}{\bibfnamefont{A.}~\bibnamefont{Valance}},
  \bibinfo{journal}{J. Math. Phys} \textbf{\bibinfo{volume}{36}},
  \bibinfo{pages}{1572} (\bibinfo{year}{1995}).

\bibitem[{\citenamefont{Prugovecki}(1986)}]{coherentprugo1}
\bibinfo{author}{\bibfnamefont{E.}~\bibnamefont{Prugovecki}},
  \emph{\bibinfo{title}{Stochastic Quantum Mechanics and Quantum Spacetime}}
  (\bibinfo{publisher}{Reidel}, \bibinfo{address}{Dordrecht},
  \bibinfo{year}{1986}).

\bibitem[{\citenamefont{Koopman}(1931)}]{kvn1}
\bibinfo{author}{\bibfnamefont{B.~O.} \bibnamefont{Koopman}},
  \bibinfo{journal}{Proc. Natl. Acad. Sci. USA} \textbf{\bibinfo{volume}{17}},
  \bibinfo{pages}{315} (\bibinfo{year}{1931}).

\bibitem[{\citenamefont{von Neumann}(1932{\natexlab{a}})}]{kvn2}
\bibinfo{author}{\bibfnamefont{J.}~\bibnamefont{von Neumann}},
  \bibinfo{journal}{Ann. Math.} \textbf{\bibinfo{volume}{33}},
  \bibinfo{pages}{587} (\bibinfo{year}{1932}{\natexlab{a}}).

\bibitem[{\citenamefont{von Neumann}(1932{\natexlab{b}})}]{kvn3}
\bibinfo{author}{\bibfnamefont{J.}~\bibnamefont{von Neumann}},
  \bibinfo{journal}{Ann. Math.} \textbf{\bibinfo{volume}{33}},
  \bibinfo{pages}{789} (\bibinfo{year}{1932}{\natexlab{b}}).

\bibitem[{\citenamefont{Gozzi et~al.}(1989)\citenamefont{Gozzi, Reuter, and
  Thacker}}]{kvn4}
\bibinfo{author}{\bibfnamefont{E.}~\bibnamefont{Gozzi}},
  \bibinfo{author}{\bibfnamefont{M.}~\bibnamefont{Reuter}}, \bibnamefont{and}
  \bibinfo{author}{\bibfnamefont{W.~D.} \bibnamefont{Thacker}},
  \bibinfo{journal}{Phys. Rev. D} \textbf{\bibinfo{volume}{40}},
  \bibinfo{pages}{3363} (\bibinfo{year}{1989}).

\bibitem[{\citenamefont{Gozzi and Reuter}(1989)}]{kvn5}
\bibinfo{author}{\bibfnamefont{E.}~\bibnamefont{Gozzi}} \bibnamefont{and}
  \bibinfo{author}{\bibfnamefont{M.}~\bibnamefont{Reuter}},
  \bibinfo{journal}{Phys. Lett. B} \textbf{\bibinfo{volume}{233}},
  \bibinfo{pages}{383} (\bibinfo{year}{1989}).

\bibitem[{\citenamefont{Mauro}(2002)}]{kvn6}
\bibinfo{author}{\bibfnamefont{D.}~\bibnamefont{Mauro}}, \bibinfo{journal}{Int.
  J. Mod. Phys. A} \textbf{\bibinfo{volume}{17}}, \bibinfo{pages}{1301}
  (\bibinfo{year}{2002}).

\bibitem[{\citenamefont{Bergeron}(2001)}]{kvnberg}
\bibinfo{author}{\bibfnamefont{H.}~\bibnamefont{Bergeron}},
  \bibinfo{journal}{J. Math. Phys.} \textbf{\bibinfo{volume}{42}},
  \bibinfo{pages}{3983} (\bibinfo{year}{2001}).

\bibitem[{\citenamefont{Mackey}(1963)}]{unifiedformalism1}
\bibinfo{author}{\bibfnamefont{G.}~\bibnamefont{Mackey}},
  \emph{\bibinfo{title}{The Mathematical Foundations of Quantum Mechanics}}
  (\bibinfo{publisher}{Benjamin}, \bibinfo{address}{New York},
  \bibinfo{year}{1963}).

\bibitem[{\citenamefont{Mackey}(1968)}]{unifiedformalism2}
\bibinfo{author}{\bibfnamefont{G.}~\bibnamefont{Mackey}},
  \emph{\bibinfo{title}{Induced Representations of Groups and Quantum
  Mechanics}} (\bibinfo{publisher}{Benjamin}, \bibinfo{address}{New York},
  \bibinfo{year}{1968}).

\bibitem[{\citenamefont{Prugovecki}(1992)}]{unifiedprugo1}
\bibinfo{author}{\bibfnamefont{E.}~\bibnamefont{Prugovecki}},
  \emph{\bibinfo{title}{Quantum Geometry}} (\bibinfo{publisher}{Kluwer},
  \bibinfo{address}{Dordrecht}, \bibinfo{year}{1992}).

\bibitem[{\citenamefont{Prugovecki}(1995)}]{unifiedprugo2}
\bibinfo{author}{\bibfnamefont{E.}~\bibnamefont{Prugovecki}},
  \emph{\bibinfo{title}{Principles of Quantum General Relativity}}
  (\bibinfo{publisher}{World Scientific}, \bibinfo{address}{Singapore},
  \bibinfo{year}{1995}).

\bibitem[{\citenamefont{Bohm}(1952)}]{bfn1}
\bibinfo{author}{\bibfnamefont{D.}~\bibnamefont{Bohm}}, \bibinfo{journal}{Phys.
  Rev.} \textbf{\bibinfo{volume}{85}}, \bibinfo{pages}{166}
  (\bibinfo{year}{1952}).

\bibitem[{\citenamefont{Bohm}(1993)}]{bfn2}
\bibinfo{author}{\bibfnamefont{D.}~\bibnamefont{Bohm}},
  \emph{\bibinfo{title}{The undivided universe: an ontological interpretation
  of quantum theory}} (\bibinfo{publisher}{Routledge and Kegan},
  \bibinfo{address}{London}, \bibinfo{year}{1993}).

\bibitem[{\citenamefont{Bohm and Vigier}(1954)}]{bfn3}
\bibinfo{author}{\bibfnamefont{D.}~\bibnamefont{Bohm}} \bibnamefont{and}
  \bibinfo{author}{\bibfnamefont{J.~P.} \bibnamefont{Vigier}},
  \bibinfo{journal}{Phys. Rev.} \textbf{\bibinfo{volume}{96}},
  \bibinfo{pages}{208} (\bibinfo{year}{1954}).

\bibitem[{\citenamefont{F\'{e}nyes}(1952)}]{bfn4}
\bibinfo{author}{\bibfnamefont{I.}~\bibnamefont{F\'{e}nyes}},
  \bibinfo{journal}{Z. Phys.} \textbf{\bibinfo{volume}{132}},
  \bibinfo{pages}{81} (\bibinfo{year}{1952}).

\bibitem[{\citenamefont{Nelson}(1966)}]{bfn5}
\bibinfo{author}{\bibfnamefont{E.}~\bibnamefont{Nelson}},
  \bibinfo{journal}{Phys. Rev.} \textbf{\bibinfo{volume}{150}},
  \bibinfo{pages}{1079} (\bibinfo{year}{1966}).

\bibitem[{\citenamefont{Nelson}(1967)}]{bfn6}
\bibinfo{author}{\bibfnamefont{E.}~\bibnamefont{Nelson}},
  \emph{\bibinfo{title}{Dynamical Theories of Brownian Motion}}
  (\bibinfo{publisher}{Princeton University Press},
  \bibinfo{address}{Princeton}, \bibinfo{year}{1967}).

\bibitem[{\citenamefont{Nelson}(1985)}]{bfn7}
\bibinfo{author}{\bibfnamefont{E.}~\bibnamefont{Nelson}},
  \emph{\bibinfo{title}{Quantum Fluctuations}} (\bibinfo{publisher}{Princeton
  University Press}, \bibinfo{address}{Princeton}, \bibinfo{year}{1985}).

\bibitem[{\citenamefont{de~la Pena-Auerbach}(1967)}]{bfn8}
\bibinfo{author}{\bibfnamefont{L.}~\bibnamefont{de~la Pena-Auerbach}},
  \bibinfo{journal}{Phys. Lett.} \textbf{\bibinfo{volume}{24A}},
  \bibinfo{pages}{603} (\bibinfo{year}{1967}).

\bibitem[{\citenamefont{de~la Pena-Auerbach}(1968)}]{bfn9}
\bibinfo{author}{\bibfnamefont{L.}~\bibnamefont{de~la Pena-Auerbach}},
  \bibinfo{journal}{Phys. Lett.} \textbf{\bibinfo{volume}{27A}},
  \bibinfo{pages}{594} (\bibinfo{year}{1968}).

\bibitem[{\citenamefont{Blanchard et~al.}(1986)\citenamefont{Blanchard, Golin,
  and Serva}}]{bfn10a}
\bibinfo{author}{\bibfnamefont{P.}~\bibnamefont{Blanchard}},
  \bibinfo{author}{\bibfnamefont{S.}~\bibnamefont{Golin}}, \bibnamefont{and}
  \bibinfo{author}{\bibfnamefont{M.}~\bibnamefont{Serva}},
  \bibinfo{journal}{Phys. Rev. D} \textbf{\bibinfo{volume}{34}},
  \bibinfo{pages}{3732} (\bibinfo{year}{1986}).

\bibitem[{\citenamefont{Bell}(1987)}]{bfn10b}
\bibinfo{author}{\bibfnamefont{J.~S.} \bibnamefont{Bell}},
  \emph{\bibinfo{title}{Speakable and Unspeakable in Quantum Mechanics}}
  (\bibinfo{publisher}{Cambridge University Press},
  \bibinfo{address}{Cambridge}, \bibinfo{year}{1987}).

\bibitem[{\citenamefont{Correggi and Morchio}(2002)}]{bfn10c}
\bibinfo{author}{\bibfnamefont{M.}~\bibnamefont{Correggi}} \bibnamefont{and}
  \bibinfo{author}{\bibfnamefont{G.}~\bibnamefont{Morchio}},
  \bibinfo{journal}{Ann. Phys.} \textbf{\bibinfo{volume}{296}},
  \bibinfo{pages}{371} (\bibinfo{year}{2002}).

\bibitem[{\citenamefont{Holland et~al.}(1986)\citenamefont{Holland,
  Kyprianidis, Maric, and Vigier}}]{bfn11}
\bibinfo{author}{\bibfnamefont{P.~R.} \bibnamefont{Holland}},
  \bibinfo{author}{\bibfnamefont{A.}~\bibnamefont{Kyprianidis}},
  \bibinfo{author}{\bibfnamefont{Z.}~\bibnamefont{Maric}}, \bibnamefont{and}
  \bibinfo{author}{\bibfnamefont{J.~P.} \bibnamefont{Vigier}},
  \bibinfo{journal}{Phys. Rev. A} \textbf{\bibinfo{volume}{33}},
  \bibinfo{pages}{4380} (\bibinfo{year}{1986}).

\bibitem[{\citenamefont{Takabayasi}(1954)}]{bfn12}
\bibinfo{author}{\bibfnamefont{T.}~\bibnamefont{Takabayasi}},
  \bibinfo{journal}{Prog. Theor. Phys.} \textbf{\bibinfo{volume}{11}},
  \bibinfo{pages}{341} (\bibinfo{year}{1954}).

\bibitem[{\citenamefont{Leavens and Mayato}(2001)}]{bfn13}
\bibinfo{author}{\bibfnamefont{C.~R.} \bibnamefont{Leavens}} \bibnamefont{and}
  \bibinfo{author}{\bibfnamefont{R.~S.} \bibnamefont{Mayato}},
  \bibinfo{journal}{Phys. Lett. A.} \textbf{\bibinfo{volume}{280}},
  \bibinfo{pages}{163} (\bibinfo{year}{2001}).

\bibitem[{\citenamefont{Polavieja}(1996)}]{bfn14}
\bibinfo{author}{\bibfnamefont{G.~G.} \bibnamefont{Polavieja}},
  \bibinfo{journal}{Phys. Lett. A} \textbf{\bibinfo{volume}{220}},
  \bibinfo{pages}{303} (\bibinfo{year}{1996}).

\bibitem[{\citenamefont{Dias and Prata}(2001)}]{bfn15}
\bibinfo{author}{\bibfnamefont{N.~C.} \bibnamefont{Dias}} \bibnamefont{and}
  \bibinfo{author}{\bibfnamefont{J.~N.} \bibnamefont{Prata}},
  \bibinfo{journal}{Phys. Lett. A} \textbf{\bibinfo{volume}{291}},
  \bibinfo{pages}{355} (\bibinfo{year}{2001}).

\bibitem[{\citenamefont{Dias and Prata}(2002)}]{bfn16}
\bibinfo{author}{\bibfnamefont{N.~C.} \bibnamefont{Dias}} \bibnamefont{and}
  \bibinfo{author}{\bibfnamefont{J.~N.} \bibnamefont{Prata}},
  \bibinfo{journal}{Phys. Lett. A} \textbf{\bibinfo{volume}{302}},
  \bibinfo{pages}{261} (\bibinfo{year}{2002}).

\bibitem[{\citenamefont{Born}(1956)}]{born}
\bibinfo{author}{\bibfnamefont{M.}~\bibnamefont{Born}},
  \emph{\bibinfo{title}{Physics in My Generation}}
  (\bibinfo{publisher}{Pergamon Press}, \bibinfo{address}{London},
  \bibinfo{year}{1956}).

\bibitem[{\citenamefont{Holland}(1993)}]{holland}
\bibinfo{author}{\bibfnamefont{P.~R.} \bibnamefont{Holland}},
  \emph{\bibinfo{title}{The Quantum Theory of Motion}}
  (\bibinfo{publisher}{Cambridge University Press},
  \bibinfo{address}{Cambridge}, \bibinfo{year}{1993}).

\bibitem[{\citenamefont{Hall and Reginatto}(2002)}]{hallreginatto}
\bibinfo{author}{\bibfnamefont{M.~J.~W.} \bibnamefont{Hall}} \bibnamefont{and}
  \bibinfo{author}{\bibfnamefont{M.}~\bibnamefont{Reginatto}},
  \bibinfo{journal}{Fortschritte der Physik} \textbf{\bibinfo{volume}{50}},
  \bibinfo{pages}{646} (\bibinfo{year}{2002}).

\bibitem[{\citenamefont{Reginatto}(1998{\natexlab{a}})}]{reginatto1}
\bibinfo{author}{\bibfnamefont{M.}~\bibnamefont{Reginatto}},
  \bibinfo{journal}{Phys. Rev. A} \textbf{\bibinfo{volume}{58}},
  \bibinfo{pages}{1775} (\bibinfo{year}{1998}{\natexlab{a}}).

\bibitem[{\citenamefont{Reginatto}(1998{\natexlab{b}})}]{reginatto2}
\bibinfo{author}{\bibfnamefont{M.}~\bibnamefont{Reginatto}},
  \bibinfo{journal}{Phys. Lett. A} \textbf{\bibinfo{volume}{249}},
  \bibinfo{pages}{355} (\bibinfo{year}{1998}{\natexlab{b}}).

\bibitem[{\citenamefont{Davidson}(1979)}]{davidson}
\bibinfo{author}{\bibfnamefont{M.}~\bibnamefont{Davidson}},
  \bibinfo{journal}{J. Math. Phys.} \textbf{\bibinfo{volume}{20}},
  \bibinfo{pages}{1865} (\bibinfo{year}{1979}).

\bibitem[{\citenamefont{Kaniadakis}(2002)}]{kaniadakis}
\bibinfo{author}{\bibfnamefont{G.}~\bibnamefont{Kaniadakis}},
  \bibinfo{journal}{Physica A} \textbf{\bibinfo{volume}{307}},
  \bibinfo{pages}{172} (\bibinfo{year}{2002}).

\bibitem[{\citenamefont{Landau and Lifchitz}(1974)}]{meca1}
\bibinfo{author}{\bibfnamefont{L.}~\bibnamefont{Landau}} \bibnamefont{and}
  \bibinfo{author}{\bibfnamefont{E.}~\bibnamefont{Lifchitz}},
  \emph{\bibinfo{title}{M\'{e}canique}} (\bibinfo{publisher}{Editions Mir},
  \bibinfo{address}{Moscou}, \bibinfo{year}{1974}).

\bibitem[{\citenamefont{Golstein}(1980)}]{meca2}
\bibinfo{author}{\bibfnamefont{H.}~\bibnamefont{Golstein}},
  \emph{\bibinfo{title}{Classical Mechanics}}
  (\bibinfo{publisher}{Addison-Wesley}, \bibinfo{address}{MA},
  \bibinfo{year}{1980}).

\end{thebibliography}

\end{document}